\documentclass[10pt]{iopart}

\expandafter\let\csname equation*\endcsname\relax

\expandafter\let\csname endequation*\endcsname\relax

\usepackage{graphicx}
\usepackage{amsfonts}
\usepackage{amssymb}
\usepackage{amsmath}
\usepackage{float}
\usepackage{color}
\usepackage{ulem}
\usepackage{bm}
\usepackage{longtable}
\usepackage{array}
\usepackage{citesort}

\newcommand*{\T}[1]{\boldsymbol{#1}}
\newcommand*{\rttensor}[1]{\boldsymbol{#1}}
\newcommand*{\rvector}[1]{\boldsymbol{#1}}

\begin{document}

\title{Multifunctional Hyperuniform Cellular Networks: Optimality, Anisotropy and Disorder}

\author{S. Torquato}

\address{Department of Chemistry, Department of Physics,
Princeton Institute for the Science and Technology of
Materials, and  Program in Applied and Computational Mathematics,
Princeton University,
Princeton, New Jersey 08544, USA}

\ead{torquato@electron.princeton.edu}

\author{D. Chen}

\address{Department of Chemistry, Princeton University,
Princeton, New Jersey 08544, USA}

\date{\today}

\begin{abstract}
Disordered hyperuniform heterogeneous materials are new, exotic 
amorphous states of matter that behave like crystals in the manner 
in which they suppress volume-fraction fluctuations at large length 
scales, and yet are statistically isotropic with no Bragg peaks. 
It has recently been shown that disordered hyperuniform dielectric 
two-dimensional cellular network solids possess complete 
photonic band gaps comparable in size 
to photonic crystals, while at the same time maintaining statistical 
isotropy, enabling waveguide geometries not possible with photonic 
crystals. Motivated by these developments, we explore other 
functionalities of various two-dimensional ordered and disordered 
hyperuniform cellular networks, including their effective thermal or
electrical conductivities and elastic moduli. We establish the 
multifunctionality of a class of such low-density networks by demonstrating 
that they maximize or virtually maximize the effective conductivities 
and elastic moduli. This is accomplished using the machinery of 
homogenization theory, including optimal bounds and
cross-property bounds, and statistical mechanics. We rigorously prove that
anisotropic networks consisting of sets of intersecting parallel 
channels in the low-density limit, 
ordered or disordered, possess optimal effective 
conductivity tensors. For a variety of different disordered networks, we show that 
when short-range and long-range order  increases, there 
is an increase in both the effective conductivity and 
elastic moduli of the network. Moreover, we demonstrate 
that the effective conductivity and elastic moduli of 
various disordered networks derived from disordered ``stealthy'' 
hyperuniform point patterns possess virtually optimal values. 
We note that the optimal
networks for conductivity are
also optimal for the 
fluid permeability associated
with slow viscous flow through
the channels as well
as the mean survival time 
associated with diffusion-controlled
reactions in the channels. 
In summary, we have identified ordered and disordered hyperuniform 
low-weight cellular networks that are multifunctional with respect
to transport (e.g., heat dissipation and fluid 
transport), mechanical and electromagnetic properties, 
which can be readily fabricated using 3D printing and lithographic technologies.

\end{abstract}


\ioptwocol

\maketitle

\section{Introduction}
Heterogeneous materials consisting of different phases are ideally suited to achieve a broad spectrum of desirable bulk physical properties by combining the best features of the constituents through the strategic spatial arrangement of the different phases \cite{To02a,Sa03a,To02b,To04,Zo12}. 
Multifunctional cellular network solids are 
commonly used in many applications 
due to their light weight and desirable transport, mechanical, optical and acoustic properties \cite{Gi99,Wa06,Va11,Be17,Su13,Su14,Zh14,Wi15,Iy15,Du04,Du05,Lu98,Hy00,Hy02}. For example, cellular solids are used as structural panels, energy adsorption devices and thermal insulators \cite{Gi99,Wa06,Va11}.

Motivated by the hyperuniformity
concept that enables a unified classification of ordered and
special disordered structures \cite{To03,Za09,To18}, 
this paper explores the 
multifunctionality of cellular
networks with varying degrees
of order (or disorder). The hyperuniformity notion
was first introduced in the context of many-particle systems more than a decade ago \cite{To03}.
Hyperuniform many-particle systems have density fluctuations that are 
anomalously suppressed at long wavelengths compared to those
in typical disordered point configurations, such as ideal gases, liquids, 
and glasses \cite{To03,Za09}. 
More precisely, 
a many-particle system
is hyperuniform if its structure
factor $S({\bf k})$ [defined in Eq. (\ref{eq_60})]
tends to zero as the wavenumber $k\equiv|\bf k|$ goes to zero (where $\bf k$ is the wavevector).
Hyperuniform systems include all perfect 
crystals and quasicrystals, and special disordered varieties \cite{To03,Za09}. Disordered hyperuniform 
many-particle systems are amorphous states of matter 
that lie between a crystal and a liquid: they behave like crystals in the way that 
they suppress density fluctuations at very large length scales, and yet they are statistically 
isotropic with no Bragg peaks. In this sense, they have a 
\textit{hidden} long-range order that is not visually apparent 
\cite{To03,Za09} (see Sec. 2 for precise definitions). 
 
The concept of hyperuniformity was generalized to two-phase materials
\cite{Za09,To16a,To16b,Ch18}. A hyperuniform two-phase medium is one in which the local volume-fraction fluctuations are suppressed 
at large length scales. More precisely, a two-phase
system is hyperuniform if its spectral density
$\widetilde{\chi}_{_V}({\bf k})$ (defined in Sec. 2)
tends to zero as $k$ goes to zero. Clearly, any network 
can be viewed as two-phase medium consisting of a ``channel'' phase distributed 
throughout some matrix or void phase. Recently, disordered hyperuniform two-phase materials
were found to possess desirable transport and mechanical properties, and wave-propagation 
characteristics \cite{Zh16,Ch18,Xu17}. 

Disordered ``stealthy'' hyperuniform dieletric two-dimensional networks \cite{Fl09,Ma13} 
are novel cellular solids that
have recently been shown to possess 
complete photonic band gaps comparable in size to photonic crystals,
while at the same time maintaining statistical isotropy, enabling waveguide geometries 
not possible with photonic crystals \cite{Fl09,Ma13}. 
Stealthy patterns are not
only hyperuniform but they possess zero-scattering
intensity for a range of wavenumbers around the origin
(see Sec. 2 for a precise definition). 
Disordered stealthy hyperuniform 
materials can be thought of
as an exotic state of matter intermediate between a crystal and a liquid \cite{To18}.
This photonic study provides a vivid example
of a class of disordered materials
that has advantages over ordered
counterparts and has led to a flurry of
papers on the study of photonic properties of disordered hyperuniform
networks \cite{Fl13,De16,Le16,Gk17,Mu17}. 
It has been suggested \cite{To18}
that the novel properties 
associated with disordered
stealthy networks is related to the fact that they cannot possess
arbitrarily large ``holes" (or cells) \cite{Zh17}. 
In addition, disordered stealthy hyperuniform two-phase materials
were recently found to possess desirable transport properties \cite{Zh16,Ch18}. 

Motivated by these developments, 
we explore other functionalities
of various two-dimensional ordered
and disordered hyperuniform cellular 
networks in the low-density limit, including
their effective thermal or electrical conductivities 
and elastic moduli. 
Our overall objective is to investigate how hyperuniformity 
affects the effective conductivity and elastic moduli of the networks, 
and how close disordered hyperuniform networks, under the 
constraint of isotropy, can come to being optimal, 
i.e., maximal with respect to these physical properties. 
We establish the multifunctionality of a class of such networks 
by demonstrating that they maximize or virtually maximize
the effective conductivities and elastic
moduli. This is accomplished
using the machinery of homogenization
theory, including optimal bounds and cross-property bounds,
and statistical mechanics. By mathematical analogy, all of our results 
for the effective conductivity apply as well to the effective dielectric constant and effective magnetic permeability \cite{To02a}. 
In addition, our results for the effective
conductivity are also optimal for the fluid permeability 
and mean survival time (see Sec. 7 for details).
 
For purposes of comparison, we first investigate the effective properties of ordered (periodic) hyperuniform networks, which include both 
macroscopically isotropic and anisotropic varieties 
\footnote{Macroscopic anisotropy refers
to an anisotropic effective property tensor. Macroscopic isotropy refers
to an isotropic effective property tensor.}. 
Then we study various disordered networks that are statistically isotropic 
derived from Voronoi, Delaunay, and what we term as 
``Delaunay-centroidal'' tessellations derived from 
hyperuniform and nonhyperuniform point patterns. 
We employ theoretical and simulation techniques, rigorous bounds, and cross-property bounds to determine the effective 
conductivity and elastic moduli of the networks. 
To quantify how close the effective conductivity tensor of 
an anisotropic network is to being optimal (i.e., maximal), we introduce
and compute the {\it tortuosity}
tensor $\rttensor{\tau}$. 

We rigorously demonstrate for the first time that anisotropic networks consisting of 
sets of intersecting parallel channels possess 
optimal effective conductivity tensors. 
It is noteworthy that this proof applies to disordered hyperuniform 
and nonhyperuniform varieties, where the 
parallel channels in each set are not equally spaced. 
We generally find that when short-range and long-range order of a Voronoi, 
Delaunay, or ``Delaunay-centroidal'' network 
increases, there is an increase in both the effective conductivity and bulk moduli of the network, and the shear moduli in the cases of Delaunay networks. Moreover, we demonstrate that the effective conductivity and bulk moduli of certain disordered networks derived from 
disordered stealthy hyperuniform 
point patterns, and the shear moduli of certain Delaunay networks possess virtually optimal values. 

The rest of the paper is organized as follows:
in Sec. 2, we provide key definitions and preliminary discussion. 
In Sec. 3, we briefly review basic results from homogenization theory that
are applied in this paper. In Sec. 4, we apply the general homogenization theory 
to low-density network solids and derive specific results for these structures.
In Sec. 5, we determine the effective conductivity and elastic moduli for various 
periodic hyperuniform networks, and compute the tortuosity tensors of 
these networks. We also provide a
rigorous proof that networks consisting of intersecting parallel channels possess optimal effective conductivity. 
In Sec. 6, we determine the effective 
conductivity and elastic moduli for various disordered hyperuniform 
and nonhyperuniform networks. In Sec. 7, we discuss
the results and provide concluding remarks.

\section{Definitions And Preliminaries}
\subsection{Point Patterns}
A statistically homogeneous point pattern in $d$-dimensional Euclidean space $\mathbb{R}^d$ 
at number density $\rho$
is characterized by its $n$-particle correlation function $g_n$ \cite{To02a}. A periodic point pattern 
represents a special subset of point patterns. It is obtained by placing a fixed 
configuration of $N$ points (where $N\geq1$) within one fundamental cell (the smallest 
repeating unit), which is then periodically replicated \cite{To10}.

Often in practice only lower-order statistics are available for statistically homogeneous point patterns. 
The pair correlation function
$g_2({\bf r})$ is a particularly
important quantity, which is defined
to be proportional to the probability of finding a point at a displacement 
of ${\bf r}$ away from a given reference point \cite{To02a}. 
The structure factor $S({\bf k})$ is essentially related
to the Fourier transform
of $g_2({\bf r})$; specifically, it is
given in terms of the Fourier transform ${\tilde h}({\bf k})$ 
of total correlation function $h({\bf r})\equiv g_2({\bf r})-1$ \cite{To02a}
via
\begin{equation}
\label{eq_60} S({\bf k})= 1+ \rho {\tilde h}({\bf k}),
\end{equation}
where ${\bf k}$ is the wave vector.

A hyperuniform many-particle system in $d$-dimensional 
Euclidean space $\mathbb{R}^d$ at number density $\rho$ 
is one in which the structure factor $S({\bf k})$ 
tends to zero as the wavenumber $k\equiv |{\bf k}|$ tends to zero \cite{To03,Za09}, i.e.,
\begin{equation}
\label{eq_58} \lim_{|{\bf k}|\rightarrow0} S({\bf k})=0.
\end{equation}
Equivalently, the local number density fluctuation $\sigma_{_{N}}^2(R)$ 
associated with a spherical window of radius $R$ of hyperuniform 
systems grows more slowly than the volume of 
that window \cite{To03}, i.e., slower than $R^d$. Stealthy systems are a special
hyperuniformity class in which the structure factor is identically zero
for a range of wavenumbers around the origin, i.e.,
\begin{equation}
\label{eq_63} S({\bf k}) = 0~~~~\textnormal{for}~~~ 0 \le |{\bf k}| < K,
\end{equation}
where the constant $K$ is the radius of the ``exclusion sphere''.
The ``stealthiness'' parameter 
\begin{equation}
\label{eq_n10} \chi = \frac{M(k)}{d(N-1)},
\end{equation} 
which is inversely proportional to the number density, gives a measure
of the relative fraction of constrained degrees of freedom
compared to the total number of degrees of freedom $d(N-1)$ 
(subtracting out the system translational degrees
of freedom) \cite{To15}. 
Here $M(k)$ is the number of independently constrained wave vectors in the exclusion region, 
and $N$ is the number of points in the system \cite{To15}.
For $0\leq\chi<1/2$, the ground states are highly degenerate and
overwhelmingly disordered \cite{Uc04,Zh15}. Moreover, short-range order 
(tendency for particles to repel one another) increases as $\chi$ increases; 
at $\chi = 1/2$, the entropically favored ground states undergo 
a transition from disordered states to
crystalline states \cite{Uc04,Zh15}.

\subsection{Two-phase Materials}
A two-phase random medium is a domain of space $V$ in $\mathbb{R}^d$ 
that is partitioned into two disjoint regions: a phase 1 region $V_1$ and 
a phase 2 region $V_2$ such that $V_1 \cup V_2 = V$ \cite{To02a}.
The microstructure of a random two-phase medium is uniquely determined by the 
indicator functions ${\cal I}^{(p)}({\bf x})$ associated with the two individual phases ($p=1,2$) 
defined as
\begin{equation}
\label{eq_61} 
{\cal I}^{(p)}({\bf x}) =
\begin{cases}
1, &{\bf x} \;\text{in phase} \;p, \\
                    0, & \text{otherwise},
\end{cases}
\end{equation}
For statistically homogeneous two-phase materials where there are no preferred centers, the two-point probability function $S_2^{(p)}({\bf r})$ 
measures the probability of finding two points separated by vector displacement ${\bf r}$ in phase $p$ \cite{To02a}. The autocovariance function $\chi_{_{V}}({\bf r})$ is trivially related to $S_2^{(p)}({\bf r})$ via
\begin{equation}
\label{eq_62} \chi_{_{V}}({\bf r}) \equiv S_2^{(p)}({\bf r}) - \phi_p^2,
\end{equation}
where $\phi_p$ is the volume fraction of phase $p$ \cite{To02a}. The spectral density 
$\widetilde{\chi}_{_V}({\bf k})$ is the Fourier transform of the autocovariance function 
$\chi_{_V}({\bf r})$, where ${\bf k}$ is the wavevector. The 
spectral density $\widetilde{\chi}_{_V}({\bf k})$ can be viewed as the counterpart 
of $S({\bf k})$ in the two-phase context. 

A hyperuniform two-phase medium in $d$-dimensional Euclidean 
space $\mathbb{R}^d$ is one in which the spectral density 
$\widetilde{\chi}_{_V}({\bf k})$
tends to zero as the wavenumber $k$ tends to zero \cite{Za09}, i.e.,
\begin{equation}
\label{eq_65} \lim_{|{\bf k}|\rightarrow0} \widetilde{\chi}_{_V}({\bf k})=0.
\end{equation}
Equivalently, the local volume-fraction fluctuation $\sigma_{_{V}}^2(R)$ 
associated with a spherical window of radius $R$ of hyperuniform 
media decay more rapidly than the inverse of the volume of window, 
i.e., faster than $R^{-d}$, while typical disordered two-phase media 
have $R^{-d}$ decay \cite{Dr15,To16a}. 
Specifically, in the case of disordered hyperuniform two-phase media, 
the spectral density $\widetilde{\chi}_{_V}({\bf k})$ tends 
to zero in the limit $|{\bf k}|\rightarrow0$ with the power-law form \cite{Za09}
\begin{equation}
\label{eq_59} \widetilde{\chi}_{_V}({\bf k}) \sim |{\bf k}|^{\gamma},
\end{equation}
where $\gamma$ is a positive exponent ($\gamma>0$). 
Note that  the magnitude of $\gamma$  provides a rough 
measure of short-range order
in the system; as $\gamma$ 
tends to infinity, the systems tend towards stealthy two-phase media 
in which $\widetilde{\chi}_{_V}({\bf k})$ is identically zero
for a range of wavenumbers around the origin, i.e., 
\begin{equation}
\label{eq_64} \widetilde{\chi}_{_V}({\bf k}) = 0~~~~\textnormal{for}~~~ 0 \le |{\bf k}| < K.
\end{equation}

\subsection{Tessellations}
\begin{figure}[ht!]
\begin{center}
$\begin{array}{c}
\includegraphics[width=0.40\textwidth]{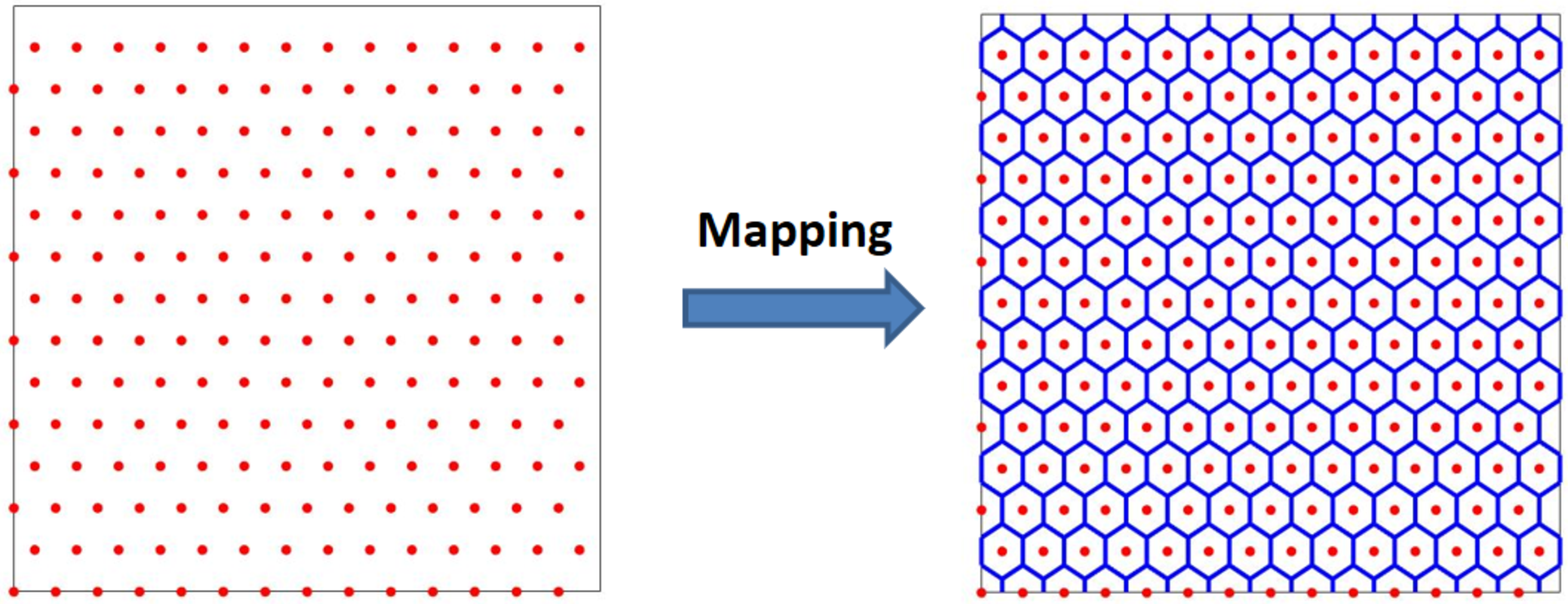} \\
\includegraphics[width=0.40\textwidth]{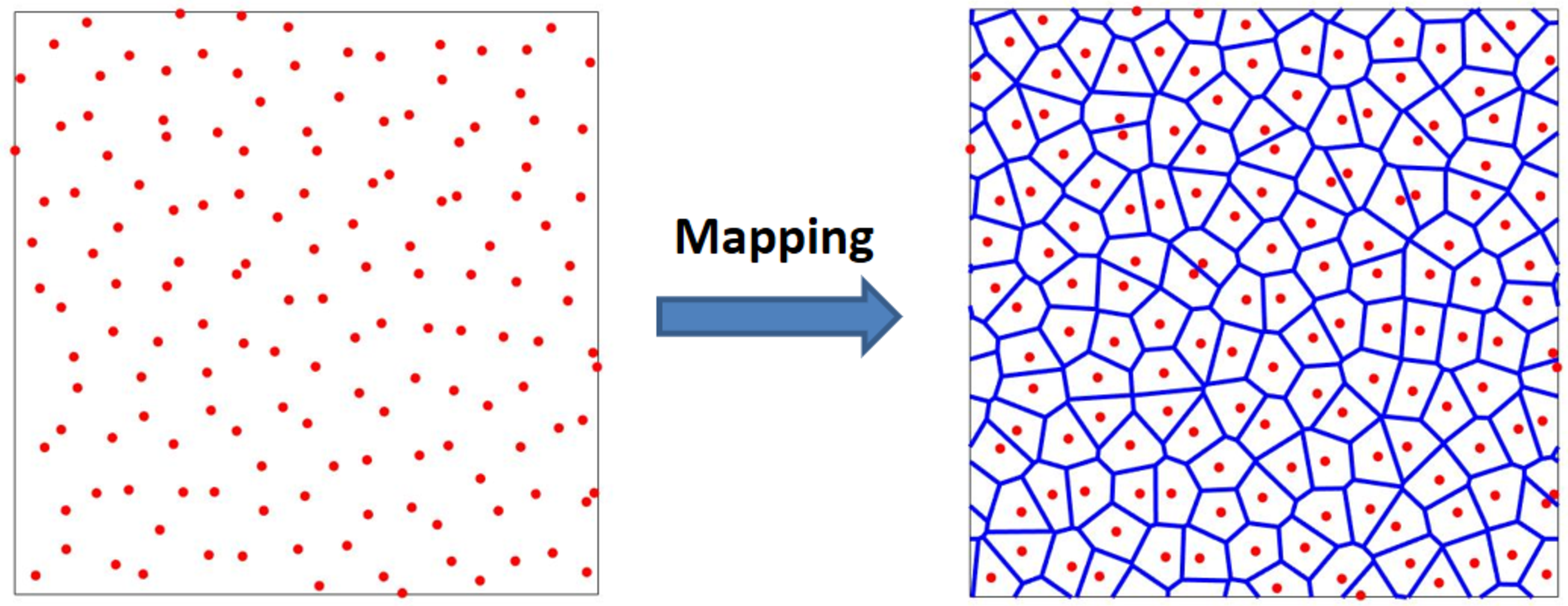}
\end{array}$
\end{center}
\caption{Schematic illustrations that demonstrate the process of mapping point patterns into networks subject to periodic boundary conditions, including one ordered example (top row) and one disordered example (bottom row). Specifically, we partition the space by using certain tessellations of 
the space based on certain point patterns 
and then decorate the edges of the resulting polygons in the tessellations with infinitely thin conducting ``channels''. In the top row, an ordered point pattern of triangular lattice is mapped into a honeycomb network via Voronoi tessellation, and in the bottom row, a disordered stealthy point pattern is mapped into a disordered network via Voronoi tessellation.}  
\label{fig_8}
\end{figure}

We map point patterns in two-dimensional Euclidean space $\mathbb{R}^2$ into two-dimensional (2D) cellular network structure by using different types of tessellations of 
the space into polygonal cells based on the underlying patterns. Then we decorate the edges of the resulting 
polygons in the tessellations with infinitely thin conducting elastic ``channels'', 
as schematically shown in Fig. \ref{fig_8}.
Specifically, we consider three types of tessellations: Delaunay, Voronoi, and 
``Delaunay-centroidal'' tessellations subject to periodic boundary 
conditions \cite{Fl09}. A
Voronoi cell is the region of space closest to a point than to any other point in the underlying patterns \cite{To02a}. A Voronoi tessellation
 is a tessellation of the space by the Voronoi cells. The Delaunay tessellation is the dual graph of the Voronoi tessellation.
The Delaunay-centroidal tessellation is generated by connecting the centroids of the neighboring triangles (which share a common edge)
in the Delaunay tessellation \cite{Fl09}. 

\section{Basic Results of Homogenization Theory}
Here we collect basic results from homogenization theory
of heterogeneous media that are central to this paper. 
This includes strong-contrast
expansions, generalized optimal 
Hashin-Shtrikman structures for
anisotropic media, rigorous
effective conductivity bounds 
and cross-property bounds
between the effective conductivity
and effective elastic moduli.
\subsection{Local and Homogenized Equations}

Consider a large two-phase system in $d$-dimensional Euclidean space $\mathbb{R}^d$ composed of two isotropic phases
with electrical (or thermal) conductivities $\sigma_1$ and $\sigma_2$. Ultimately,
we will take the infinite-volume limit. The local scalar conductivity ${\sigma}({\bf x})$ at position $\bf x$ is expressible as
\begin{equation}
{\sigma}({\bf x}) = \sigma_1{\cal I}^{(1)}({\bf x})+ \sigma_2{\cal I}^{(2)}({\bf x}) ,
\label{eq_1}
\end{equation}
where ${\cal I}^{(p)}({\bf x})$
is the indicator function for phase $p$ ($p=1,2$) defined in Eq. (\ref{eq_61}). 
The local constitutive relation, Ohm's law
in the case of electrical conduction or Fourier's law in the case of thermal conduction, is given by
\begin{equation}
\label{eq_3} \rvector{J}({\bf x})=\sigma({\bf x})\rvector{E}({\bf x}),
\end{equation}
where $\rvector{J}({\bf x})$ and $\rvector{E}({\bf x})$ denote the local flux vector
and field (equal to the negative of the gradient of the potential), respectively.
Under steady-state conditions, the local flux and field  respectively
satisfy the divergence-free and curl-free relations:
\begin{equation}
\label{eq_4} \boldsymbol{\triangledown} \cdot \rvector{J}({\bf x})=0.
\end{equation}
\begin{equation}
\label{eq_5}  \boldsymbol{\triangledown} \times  \rvector{E}({\bf x})=0.
\end{equation}
Using homogenization theory \cite{To02a,Mi02}, it can be shown that
the effective electric (or thermal) conductivity second-rank tensor $\rttensor{\sigma}_e$
is determined by the averaged Ohm's (or Fourier's) law:
\begin{equation}
\label{eq_6} \langle\rvector{J}({\bf x})\rangle=\rttensor{\sigma}_e\langle\rvector{E}({\bf x})\rangle,
\end{equation}
where angular brackets denote an ensemble average,
$\langle\rvector{J}({\bf x})\rangle$ is the average flux
and $\langle\rvector{E}({\bf x})\rangle$ is the average field.

\subsection{Exact Contrast Expansions}

Consider a macroscopically anisotropic composite medium consisting
of two isotropic phases with conductivities $\sigma_p$ and $\sigma_q$ ($p \ne q$ with $p=1,2$, $q=1,2$)
that is characterized by an effective conductivity tensor ${\T{\sigma}}_e$.
A ``strong-contrast'' expansion for $\T{\sigma}_e$
was derived in Ref. \cite{Se89} that perturbs around the
generalized optimal Hashin-Shtrikman structures for anisotropic media \cite{To02a}:

\begin{equation}
\beta^{2}_{pq}\phi^{2}_p\{{\T{\sigma}}_e-{\sigma}_q {\bf I}\}^{-1} \cdot
\{{\T{\sigma}}_e+(d-1){\sigma}_q {\bf I}\} = \phi_p \beta_{pq}{\bf I}
-\sum_{n=2}^\infty \; {\bf A}_n^{(p)} \beta_{pq}^n,
\label{eq_7}
\end{equation}
where the $n$-point tensor coefficients ${\bf A}_n^{(p)}$ are certain integrals
over the $S_n^{(p)}$ associated with phase $p$ and $\bf I$ is the identity tensor
and
\begin{equation}
\beta_{pq}=\frac{\sigma_p -\sigma_q}{\sigma_p+(d-1)\sigma_q}.
\label{eq_8}
\end{equation}
For $n=2$,
\begin{equation}
{\bf A}_2^{(p)} =
\frac{d}{\Omega(d)}\int_\epsilon d2 \;{\bf t}(1,2) \left[S_{2}^{(p)}(1,2)-\phi_{p}^2\right] ,
\label{eq_9}
\end{equation}
and for $n\ge3$,
\begin{equation}
\begin{split}
{\bf A}_n^{(p)} = & \Big(\frac{-1}{\phi_p}\Big)^{n-2}
\Big(\frac{d}{\Omega(d)}\Big)^{n-1}
\int d2\cdots \int dn \\
& {\bf t}(1,2) \cdot{\bf t}(2,3)\cdots{\bf t}(n-1, n)
 \Delta_n^{(p)} (1,\ldots,n),
 \end{split}
\label{eq_10}
\end{equation}
where
\begin{equation}
{\bf t}({\bf r})= \frac{d {\bf n}{\bf n}-{\bf I}}{r^d}
\label{eq_11}
\end{equation}
is the dipole-dipole tensor,
\begin{equation}
\Omega(d)=\frac{d \pi^{d/2}}{\Gamma(1+d/2)}
\label{eq_12}
\end{equation}
is the total solid angle contained in a $d$-dimensional sphere, and

\renewcommand{\arraystretch}{1.5}
\begin{equation}
\footnotesize
\Delta^{(p)}_{\;n} = \left|
\begin{array}{cccccc}
S^{(p)}_{2}(1,2)&S^{(p)}_{1}(2) &  \cdots &  0 \\
S^{(p)}_{3}(1,2,3)&S^{(p)}_{2}(2,3)& \cdots &  0 \\
\vdots & \vdots & \ddots & \vdots  \\
S^{(p)}_{n}(1,\ldots,n)&S^{(p)}_{n-1}(2,\ldots,n)&
\cdots & S^{(p)}_{2}(n-1,n)
\end{array}
\right|
\label{eq_13}
\end{equation}
is a position-dependent determinant associated with phase $p$.

Central to this paper is the two-point tensor parameter
${\bf A}_2^{(p)}$, which we note generally
does not vanish for statistically anisotropic media, since the
two-point function $S_2^{(p)}({\bf r})$ depends on the distance $r=|\bf r|$ as well as
the orientation of the vector $\bf r$. Second, it is the only tensor parameter in
expansion (\ref{eq_7}) that is independent of the phase $p$, and hence we define
\begin{equation}
{\bf A} \equiv {\bf A_2}^{(1)}={\bf A_2}^{(2)}
\label{eq_14}
\end{equation}
Third, we observe that for macroscopically isotropic media,
\begin{equation}
{\bf A}={\bf 0},
\label{eq_15}
\end{equation}
since ${\bf A}$ is traceless, i.e., $\mbox{Tr}{\bf A}=0$. It is noteworthy
that the two-point tensor $\bf A$
also arises in strong-contrast
expansions for the effective 
stiffness tensor \cite{To97}.

\subsection{Rigorous Bounds and Optimality}

\begin{figure}[ht!]
\begin{center}
$\begin{array}{c}
\includegraphics[width=0.35\textwidth]{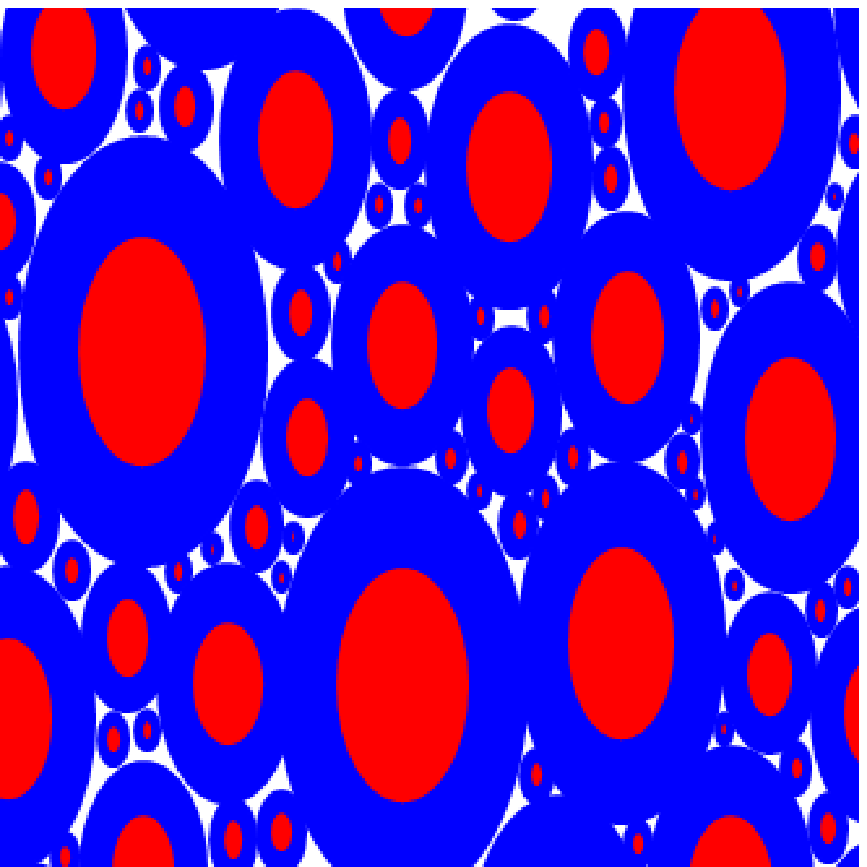} \\
\includegraphics[width=0.35\textwidth]{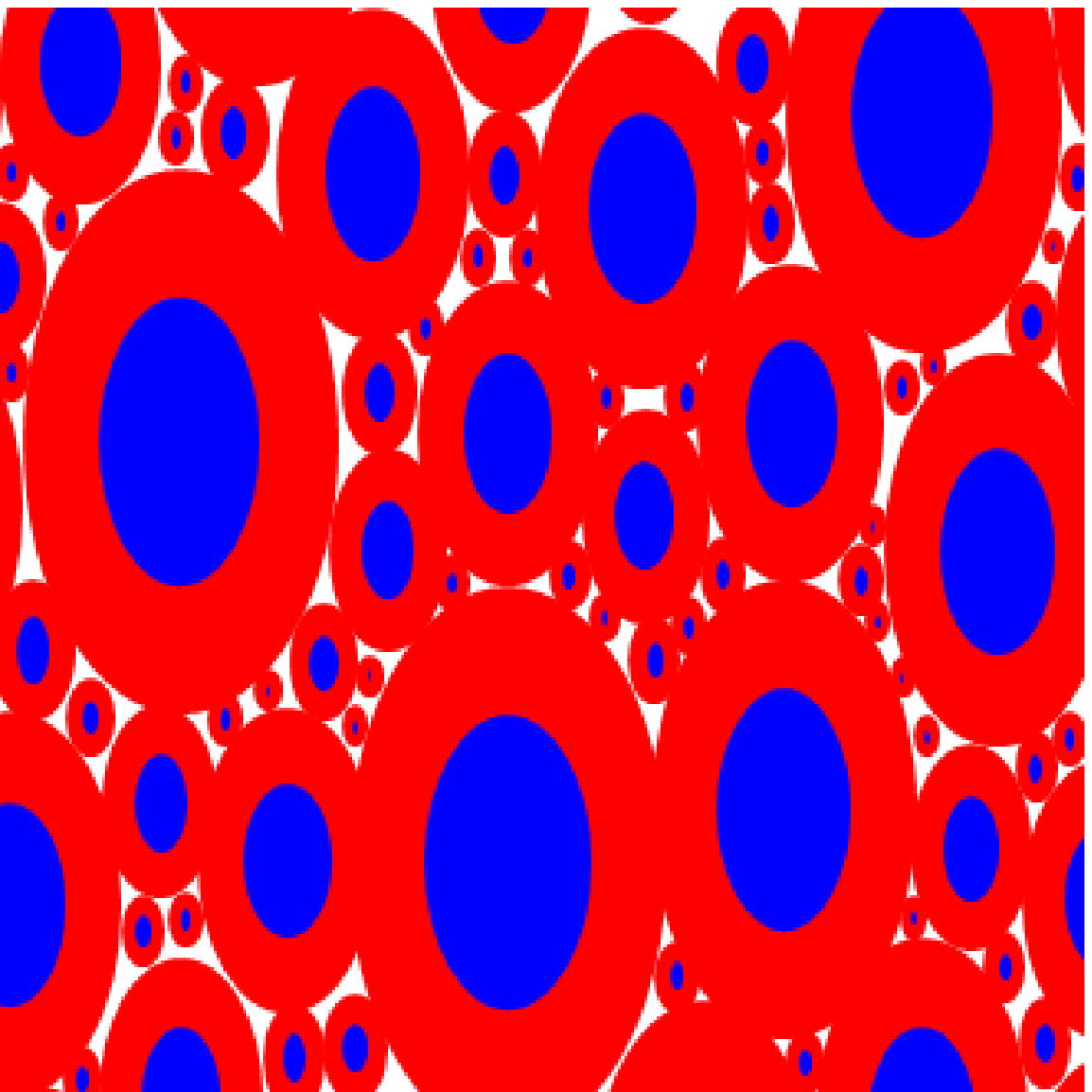}
\end{array}$
\end{center}
\caption{Coated-ellipsoid assemblages
consisting of {\it oriented} composite ellipsoids that are composed of a
ellipsoidal core of  one phase that is surrounded by a
concentric ellipsoidal shell of the other phase such that the fraction
of space occupied by the core phase is equal  to its overall phase volume fraction \cite{Be80,Mi81b,Ta85}.
The aspect ratio of the composite ellipsoid is set by the eigenvalues of the
two-point tensor ${\bf a}_2$.
The composite ellipsoids fill all space, implying that there is a distribution in their
sizes ranging to the infinitesimally small. Assuming that the red region constitutes
the more conducting phase and the blue region constitutes the less conducting phase,
the top and bottom panels show the microstructures that are exactly achieved
by the lower bound (\ref{eq_18}) and upper bound (\ref{eq_19}), respectively.}  
\label{fig_1}
\end{figure}

Rigorous bounds on the effective conductivity tensor that incorporate up to $n$-point
correlation functions are referred to as $n$-point bounds \cite{To02a}. The following
are two-point anisotropic generalizations of the Hashin--Shtrikman bounds
on $\T{\sigma}_e$ when $\sigma_2 \ge \sigma_1$:
\begin{equation}
{\T{\sigma}}_{L}^{(2)} \leq {\T{\sigma}}_{e} \leq {\T{\sigma}}_{U}^{(2)},
\label{eq_17}
\end{equation}
where
\begin{equation}
{\T{\sigma}}_{L}^{(2)} = \langle{\T{\sigma}}\rangle  + ({\sigma}_2 - {\sigma}_1)^2 {\bf a}_2 \cdot
\left[ {\sigma}_1 {\T{I}} + \frac{({\sigma}_1 - {\sigma}_2)}{\phi_2} {\bf a}_2
\right]^{-1},
\label{eq_18}
\end{equation}
\begin{equation}
{\T{\sigma}}_{U}^{(2)} = \langle {\T{\sigma}}\rangle  + ({\sigma}_2 - {\sigma}_1)^2 {\bf a}_2 \cdot
\left[ {\sigma}_2 {\T{I}} + \frac{({\sigma}_2 - {\sigma}_1)}{\phi_1} {\bf a}_2
\right]^{-1},
\label{eq_19}
\end{equation}
and
\begin{equation}
{\bf a}_2 = \frac{1}{d} \left[{\bf A} - \phi_1 \phi_2 {\bf I}\right]
\label{eq_20}
\end{equation}
is a two-point tensor parameter, which arises
in the so-called ``weak-contrast" expansion for $\T{\sigma}_e$ \cite{Se89}
and is seen to be trivially related to ${\bf A}$ and hence
obeys the trace condition
\begin{equation}
\mbox{Tr}\;{\bf a}_2= -\phi_1\phi_2.
\label{eq_21}
\end{equation}
These two-point upper and lower bounds have been derived by a variety of methods.
Willis \cite{Wi77} first derived them for $d=3$ using the anisotropic generalizations of the
Hashin--Shtrikman principles. Sen and Torquato \cite{Se89} obtained
them in arbitrary dimension $d$ using the method of Pad{\'e} approximants
\cite{Mi85a}. Importantly, the bounds (\ref{eq_18}) and (\ref{eq_19}) are achieved by
certain oriented singly-coated space-filling ellipsoidal assemblages \cite{Be80,Mi81b,Ta85} (see Fig. \ref{fig_1})
as well as finite-rank laminates \cite{Ta85}. Hence, these bounds are optimal given the phase
volume fraction and the two-point information contained in ${\bf a}_2$. For all optimal structures,
one of the phases is generally a disconnected, dispersed phase in a connected matrix phase, except in the trivial instance
in which the dispersed phase fills all of space. These two-point  bounds  are exact
to second order in the phase conductivity difference, i.e.,  for $p \neq q$, we have
\begin{equation}
{\T{\sigma}}_e= \sigma_q {\bf I} + \phi_p (\sigma_p-\sigma_q) {\bf I}+
\frac{(\sigma_p-\sigma_q)^2}{\sigma_q} {\bf a}_2  + {\cal O}\Big(\Big[ \frac{\sigma_p-\sigma_q}{\sigma_q} \Big]^3\Big).
\label{eq_22}
\end{equation}

For statistically anisotropic microstructures
in which
$S_{2}^{(p)}({\bf r})$ possesses  ellipsoidal symmetry (e.g.,
oriented similar ellipsoidal inclusions in a matrix with
nematic-liquid-crystal structure),  the aforementioned two-point
parameters are given by
\begin{equation}
{\bf A}= ({\bf I}- d {\bf A}^*)\phi_1\phi_2, \qquad {\bf a}_2=-\phi_1\phi_2{\bf A}^*,
\label{eq_23}
\end{equation}
where ${\bf A}^*$ is  the symmetric {\it depolarization} tensor
of a $d$-dimensional ellipsoid, which in the principal axes frame has diagonal components
or eigenvalues (denoted by $A^{*}_{i}$, $i=1,\ldots,d$, no summation implied) 
given by the elliptic integrals
\begin{equation}
 A^{*}_{i}=\Big(\prod_{j=1}^{d} \frac{a_j}{2} \Big)
\int_{0}^{\infty}\frac{dt}{(t+a^2_i)\sqrt{\prod_{j=1}^{d}(t+a^2_j)}}, \quad i=1,\ldots,d,
\label{eq_24}
\end{equation}
where $a_i$ is the semiaxis of the ellipsoid along the $x_i$ direction.
The depolarization tensor has the property that its trace is unity, i.e.,
\begin{equation}
\mbox{Tr}\;{\bf A}^{*}=1.
\label{eq_25}
\end{equation}
In the 2D case (ellipse) of aspect ratio $\alpha=a_2/a_1$, (\ref{eq_24}) can be simplified to yield the exact relation
\begin{equation}
{\bf A}^{*}=\left[\begin{array}{cc}
\frac{\displaystyle \alpha}{\displaystyle 1+\alpha} & 0  \\
0 & \frac{\displaystyle 1}{\displaystyle 1+\alpha}  \\
\end{array}
\right].
\label{eq_26}
\end{equation}
From these results, we see that for circles ($\alpha=1$)
\begin{equation}
A^{*}_{11}= A^{*}_{22}=\frac{1}{2},\qquad (\mbox{circles})
\label{eq_27}
\end{equation}
for needle-shaped inclusions aligned along the $x_2$-axes ($\alpha=\infty$)
and $x_1$-axes ($\alpha=0$), respectively, we have
\begin{equation}
 A^{*}_{11}=1,\quad  A^{*}_{22}=0, \qquad
(\mbox{needles along the $x_2$-axes})
\label{eq_28}
\end{equation}
and
\begin{equation}
 A^{*}_{11}=0,\quad  A^{*}_{22}=1, \qquad
(\mbox{needles along the $x_1$-axes}).
\label{eq_29}
\end{equation}

It is noteworthy, but not surprising in light of the aforementioned results, that the lower bound (\ref{eq_18}) is  exact for a dilute concentration of oriented ellipsoids ($\phi_2 \ll 1$) in a matrix of phase 1, i.e.,
\begin{equation}
{\T{\sigma}}_e =  \sigma_1 {\bf I}  + (\sigma_2 - \sigma_1) {\bf I} \cdot \Big[{\bf I}+ \frac{(\sigma_2 - \sigma_1)}{\sigma_1} {\bf A}^*\Big]^{-1} \;\phi_2 + {\cal O}(\phi_2^2).
\label{eq_30}
\end{equation}
This relation applies for any size distribution of the ellipsoids, i.e., it is not limited to identical
ellipsoids.

Whenever the two-phase system is macroscopically isotropic, i.e., ${\T{\sigma}}_e=\sigma_e {\bf I}$ and ${\bf a}_2=-\phi_1\phi_2{\bf I}/d$, where $\sigma_e$
is a scalar quantity, the two-point anisotropic bounds (\ref{eq_18}) and (\ref{eq_19})
reduce to the $d$-dimensional Hashin--Shtrikman bounds on $\sigma_e$
for two-phase isotropic media with $\sigma_2 \ge \sigma_1$:
\begin{equation}
\sigma_{L}^{(2)} \leq \sigma_{e} \leq \sigma_{U}^{(2)},
\label{eq_31}
\end{equation}
where
\begin{eqnarray}
\sigma_{L}^{(2)}&=&\langle \sigma \rangle -\frac{\phi_{1}\phi_{2}(\sigma_2-\sigma_1)^2}
{d \sigma_1 + (\sigma_2 -\sigma_1) \phi_1},
\label{eq_32}\\
\sigma_{U}^{(2)}&=&\langle \sigma \rangle -\frac{\phi_{1}\phi_{2}(\sigma_2-\sigma_1)^2}
{d \sigma_2 + (\sigma_1 -\sigma_2) \phi_2}.
\label{eq_33}
\end{eqnarray}
The Hashin--Shtrikman bounds are realized by the singly coated
$d$-dimensional sphere assemblages \cite{To02a,Mi02}, second-rank
laminates \cite{To02a,Mi02}, and single-scale Vigdergauz
constructions \cite{Vig89,Vig94}.  Accordingly, because
the bounds are attainable by certain microstructures,
they are the best possible bounds on the effective
conductivity of macroscopically isotropic two-phase composites
given volume-fraction information only.

\subsection{Cross-Property Conductivity-Elastic Moduli Bounds}

For macroscopically isotropic two-phase media, Gibiansky and Torquato \cite{Gi93,Gi96b} derived
rigorous cross-property bounds that relate the effective elastic moduli
to the effective conductivity. In the special case of two-phase media consisting of pores 
or cracks of arbitrary shape and size distributed throughout a solid material, these formulas simplify considerably \cite{Gi93,Gi96b}. Let the bulk modulus, shear modulus and Young's modulus
of the solid phase be denoted by $K$, $G$ and $E$, respectively. Denote by $K_e$, $G_e$
and $E_e$ the effective bulk modulus, shear modulus and Young's modulus, respectively. 
Note that for macroscopically isotropic structure, there are only two independent elastic 
moduli. For example, given $K_e$ and $G_e$ of a structure, any other quantities such as $E_e$
can be derived.  
The general cross-property bounds that rigorously link the
the effective elastic moduli to the effective conductivity
in 2D are given by
\begin{equation}
\frac{K}{K_e} - 1  \geq  \frac{K + G}{2 G} \left[\frac{\sigma}
{\sigma_e} - 1 \right] ,
\label{eq_37}
\end{equation}
\begin{equation}
\frac{G}{G_e} - 1  \geq  \frac{K + G}{ K} \left[\frac{\sigma}
{\sigma_e} - 1 \right] ,
\label{eq_38}
\end{equation}
\begin{equation}
\frac{E}{E_e} - 1  \geq  \frac{3}{2 } \left[\frac{\sigma}
{\sigma_e} - 1 \right] .
\label{eq_39}
\end{equation}

In the low-density asymptotic limit, i.e., $\phi \ll 1$, one can assume that $
\sigma_e/\sigma\ll 1$, $ K_e/K \ll  1$, $ G_e/G\ll  1$, and $ E_e/E \ll  1$. Under such
conditions, the cross-property bounds (\ref{eq_37})-(\ref{eq_39}) reduce to
\begin{equation}
\frac{K_e}{K} \leq \frac{2G}{K + G} \frac{\sigma_e}{\sigma} \; = (1-\nu)\frac{\sigma_e}{\sigma} \;,
\label{eq_40}
\end{equation}
\begin{equation}
\frac{G_e}{G} \leq \frac{K}{K + G} \frac{\sigma_e}{\sigma} \; = \frac{(1+\nu)}{2}\frac{\sigma_e}{\sigma} \;,
\label{eq_41}
\end{equation}
and
\begin{equation}
\frac{E_e}{E} \leq \frac{2}{3} \frac{\sigma_e}{\sigma} \;,
\label{eq_42}
\end{equation}
respectively.
Here $\nu$ is the Poisson's ratio and in 2D is bounded according to 
\begin{equation}
-1\leq\nu\leq 1 \;.
\label{eq_43}
\end{equation}

Equations (\ref{eq_37}) - (\ref{eq_42}) only apply to statistically isotropic structures or statistically anisotropic structures with 3- or 6-fold rotational symmetry. 
Note that measurement of the elastic moduli in conjunction with the cross-property
bounds (\ref{eq_37}) - (\ref{eq_42}) allows one to obtain a {\it lower bound} on the effective conductivity.
Similarly, conductivity information and bounds (\ref{eq_37}) - (\ref{eq_42}) enables one to
bound the elastic moduli from above. However, effective 
shear modulus $G_e$ and effective Young's modulus $E_e$ of certain networks, such as honeycomb-like (e.g., Voronoi and Delaunay-centroidal networks) and square-like ones, are far from optimal, i.e., 
far from the corresponding upper bounds (\ref{eq_38}), (\ref{eq_39}), (\ref{eq_42}), and (\ref{eq_43}) due to the bending modes of the structures \cite{To98a,Hy00,Hy02}. Subsequently, we will only employ the upper bounds to estimate $G_e$ and $E_e$ for the triangular networks. 

\section{Application to Low-Density Network Solids}

Of particular interest in this paper are applications of the two-point anisotropic
bounds (\ref{eq_18}) and (\ref{eq_19}) to low-density networks. Assuming
that phase 2 is the low-density, more conducting  phase (i.e., $\phi_2 \ll 1$ and $\sigma_2 \ge \sigma_1$),
these bounds become

\begin{equation}
{\T{\sigma}}_e \ge  \sigma_1 {\bf I}  + (\sigma_2 - \sigma_1) {\bf I} \cdot \Big[{\bf I}- \frac{(\sigma_1 - \sigma_2)}{\sigma_1} {\bf A}^*\Big]^{-1} \;\phi_2
\label{eq_34}
\end{equation}
\begin{equation}
{\T{\sigma}}_e \le \sigma_1 {\bf I}  + (\sigma_2 - \sigma_1) \Big[{\bf I}- \frac{(\sigma_2 - \sigma_1)}{\sigma_2} {\bf A}^*\Big] \;\phi_2
\label{eq_35}
\end{equation}
where ${\bf A}^*$ is defined in Eq. (\ref{eq_24}).

Importantly, we will see that there are anisotropic network structures in two and three dimensions
that attain the upper bound (\ref{eq_19}) and hence are optimal.
Elsewhere it was shown that macroscopically isotropic 2D ordered networks with
4-fold rotational symmetry (e.g., square tessellation) and 6-fold
rotational symmetry (e.g., honeycomb and equilateral-triangular
tessellations) attain the upper bound and hence are optimal \cite{To98a}. In these
instances, ${\bf A}^*=\phi_1\phi_2/2$, and the upper bound reduces to the corresponding
Hashin-Shtrikman upper bound, as obtained from (\ref{eq_33}).

In the extreme case in which $\sigma_1=0$,  upper bound (\ref{eq_19}) reduces to the following simple form:
\begin{equation}
{\T{\sigma}}_e \le  \rttensor{\sigma_{U}}^{(2)}=\frac{1}{2}(1+\frac{\rttensor{A}}{\phi})\phi\sigma
\label{eq_36}
\end{equation}
Henceforth when referring to the properties of
the solid phase, we drop the subscripts so that 
$\phi\equiv\phi_2$ and $\sigma\equiv\sigma_2$.
Note that the lower bound (\ref{eq_35}) is trivially zero because it corresponds to a microstructure
in which the perfectly insulting phase 1 is connected (see Fig. \ref{fig_1}).
For macroscopically isotropic media, the two-point tensor coefficient $\rttensor{A}$ 
vanishes, as stated in Eq. (\ref{eq_15}),
and the upper bound (\ref{eq_36}) 
reduces to the Hashin-Shtrikman bound $\sigma_{U}^{(2)}$ on the 
scalar effective conductivity $\sigma_e$:
\begin{equation}
\label{eq_53} \sigma_{U}^{(2)}=\frac{1}{2}\phi\sigma.
\end{equation}
In the subsequent sections, we will focus on this extreme case.

\section{Network Analysis}

In this section, we develop a general scheme to compute the effective conductivity tensor $\rttensor{\sigma}_e$
of 2D ordered and disordered cellular network
structures in which one phase consists of connected infinitesimally thin channels (henceforth called
the ``channel" phase) and the other is a disconnected and insulating ``void" phase.
We also exactly evaluate the two-point tensor $\bf A$, defined by (\ref{eq_14}), for a certain
class of such networks.

\subsection{Effective Conductivity Tensor}
Here we denote the conductivity and volume fraction of the ``channel'' 
phase by $\sigma$ and $\phi$, respectively.
To determine the effective conductivity $\rttensor{\sigma}_e$, we consider the
conduction problem in a fundamental
cell (i.e., smallest periodic repeat unit). For our purposes,
we consider rectangular fundamental cells. We set the potentials (or temperatures) at the two opposing boundaries in the $x_i$ direction to be $T_A$ and $T_B$,
and the applied field ${\bf E}_0$, which is equal to $\langle E_i \rangle$ (the average of local field in the $x_i$
direction), is given by
\begin{equation}
\label{eq_44} E_0=\langle E_i \rangle=-\frac{T_B-T_A}{L_i},
\end{equation}
where $E_0 =|{\bf E}_0|$, and $L_i$ is the side length of 
the fundamental cell in the $x_i$ direction.
In the orthogonal direction, we apply periodic boundary conditions. We denote the magnitude of the flux by $J\equiv |{\bf J}|$. 
For example, if we consider the conduction problem 
in the $x_1$ direction, the 
boundary conditions are given by
\begin{equation}
\label{eq_n1}
\left.
\begin{array}{c}
T(x_1, x_2+L_2) = T(x_1, x_2) \\
T(x_1+L_1, x_2) = T(x_1,x_2)-E_0 L_1
\end{array}
\right\}
\end{equation}
As a general guideline, when considering an applied field in one of the
orthogonal  directions, it is convenient to choose the fundamental cell so
that it possesses reflection symmetry with respect to this direction, if possible.

\begin{figure}[H]
\begin{center}
$\begin{array}{c}
\includegraphics[width=0.45\textwidth]{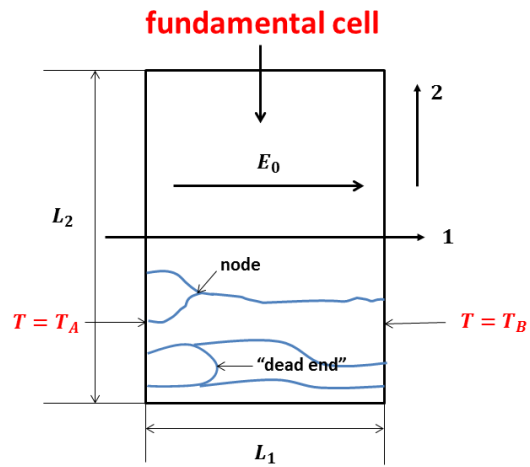}
\end{array}$
\end{center}
\caption{A general framework of the conduction problem in a fundamental cell (which in this case is in the $x_1$
direction). The lengths of the fundamental cell in the $x_1$ and $x_2$ directions are denoted by $L_1$ and $L_2$,
respectively. The potentials (or temperatures) at the two opposing boundaries in the $x_1$ direction are set to be
$T_A$ and $T_B$, and the corresponding applied field in the $x_1$ direction is denoted by $E_0$.}
\label{fig_2}
\end{figure}

Figure \ref{fig_2}
schematically shows the general setup for the conduction problem in a fundamental cell,
which, for purposes of illustration, show an applied field  ${\bf E}_0$ in
the $x_1$-direction.
The lengths of the fundamental cell in the $x_1$- and $x_2$-directions are denoted by $L_1$ and $L_2$, respectively.

The effective conductivity tensor is determined by the averaged Ohm's (Fourier's) law given by
relation (\ref{eq_6}). Since our coordinate system is aligned with the principal axes frame, then
we need only consider the diagonal components of the effective conductivity tensor.
We denote by $(\sigma_e)_{II}$ the $II$-component of the effective conductivity tensor (no summation implied).
Thus, according to Eqs. (\ref{eq_6}) and (\ref{eq_44}), we have
\begin{equation}
\label{eq_45} \langle J_i \rangle=(\sigma_e)_{II}\langle E_i \rangle=-(\sigma_e)_{II}\frac{T_B-T_A}{L_i}.
\end{equation}
where
\begin{equation}
\label{eq_46} \langle \rvector{J} \rangle=\frac{1}{\Omega}\int \rvector{J}(\rvector{x})~dV=\frac{1}{\Omega}\int
\rvector{J}(\rvector{x})~dV_2,
\end{equation}
$\Omega=L_1L_2$ is the volume of the fundamental cell, and $\int dV_2$ denotes the integral over the space occupied by the channel phase.
Applying Eq. (\ref{eq_3}) along the conduction path between opposing boundaries in the $x_i$-direction, we get
\begin{equation}
\label{eq_47} \int_{A}^{B}J~dl=-\sigma (T_B-T_A),
\end{equation}
The path integral Eq. (\ref{eq_47}) is the same for any path connecting the two opposing boundaries, which should not contain ``dead ends'', defined to be channels that are not topologically connected to boundaries 
or channels with zero flux that are not perpendicular to the applied field.
From Eqs. (\ref{eq_45}) and (\ref{eq_47}), we get
\begin{equation}
\label{eq_48} (\sigma_e)_{II}E_0=\frac{(\sigma_e)_{II}}{\sigma}\frac{\int_{A}^{B}J~dl}{L_i}=\langle J_i \rangle,
\end{equation}
Therefore, $(\sigma_e)_{II}$ is
\begin{equation}
\label{eq_49} \frac{(\sigma_e)_{II}}{\sigma}=\frac{\langle J_i \rangle L_i}{\int_{A}^{B}J~dl}.
\end{equation}
Note that in the limit $\phi\rightarrow 0$ (i.e., the thickness of the channels goes to 0), the magnitude of flux $J$ is piecewise constant for such cellular network structures. Furthermore, for structures consisting of piecewise straight channels, the flux
$\rvector{J}$ is piecewise constant. For any cellular network, letting $J_{m,n}$ represent the (signed) flux flowing
from node $m$ to $n$ (where the concept of a node schematically shown in Figure \ref{fig_1}), Ohm's and Kirchhoff's laws
can be written in a discrete form,
\begin{equation}
\label{eq_50} J_{m,n} = \sigma(T_m - T_n) H_{m,n},
\end{equation}
and similarly, the divergence-free condition (\ref{eq_4}) can be written as
\begin{equation}
\label{eq_51} \sum\limits_n J_{m, n} = 0~~~\forall m.
\end{equation}
Here $H_{m,n}$ is the generalized adjacency matrix of the graph formed by the cellular network. $H_{m,n}$ takes the
value $1.0 / a_{m,n}$, where $a_{m, n}$ is the length of the channel connecting nodes $m$ and $n$ if there exists such
a channel, and 0 otherwise. By solving Eqs. (\ref{eq_50}) and (\ref{eq_51}) and taking into account the
symmetry of the cellular network structure, we obtain the magnitude of the flux in each channel within the fundamental cell,
which is then used to compute $\langle J_i \rangle$ and $\int_{A}^{B}J~dl$. Finally, the $II$-component of the effective
conductivity $(\sigma_e)_{II}$ tensor is determined from Eq. (\ref{eq_49}).

\begin{figure*}[ht!]
\begin{center}
$\begin{array}{c@{\hspace{0.1cm}}c@{\hspace{0.1cm}}c}\\
\includegraphics[width=0.25\textwidth]{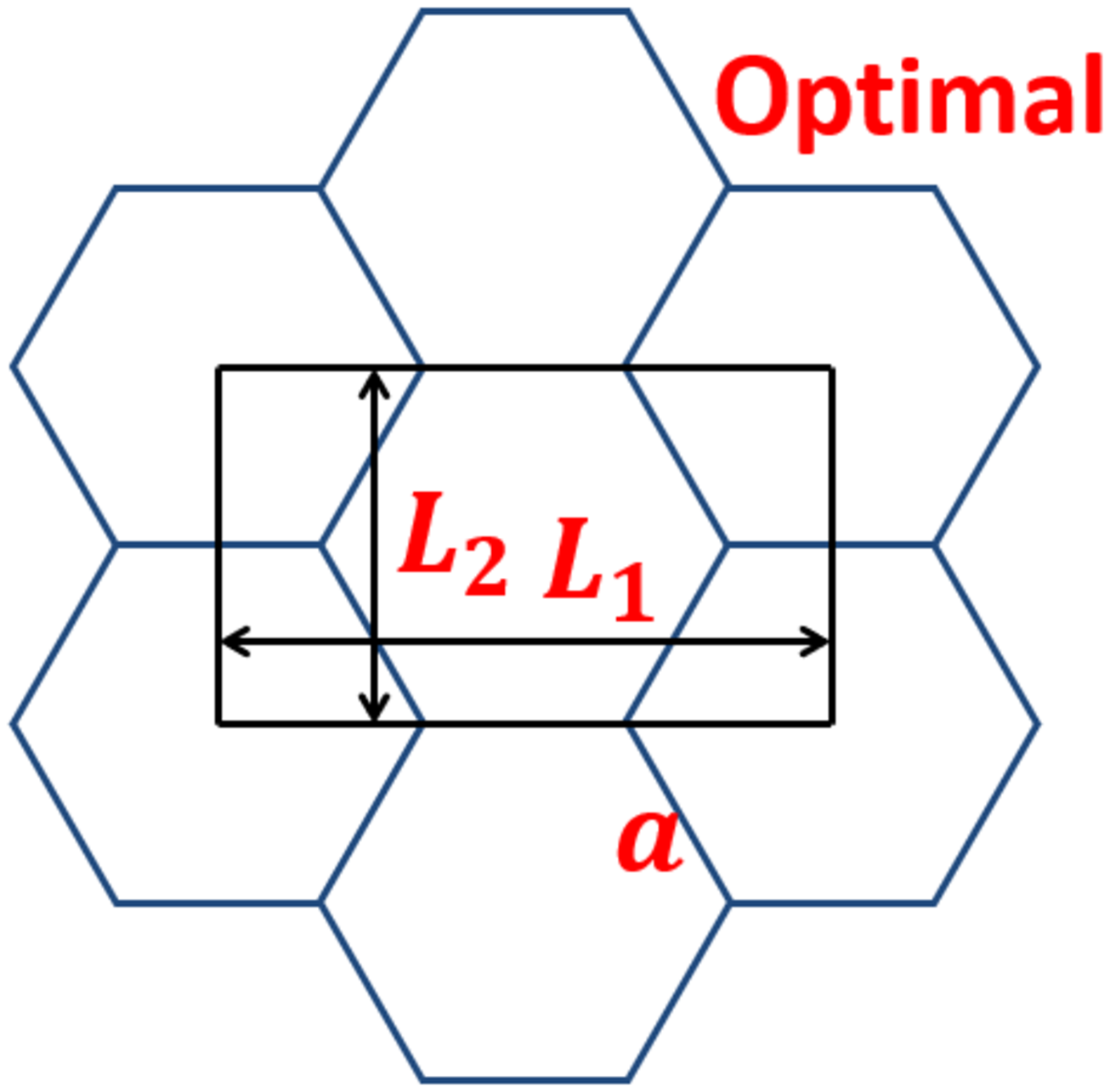} &
\includegraphics[width=0.20\textwidth]{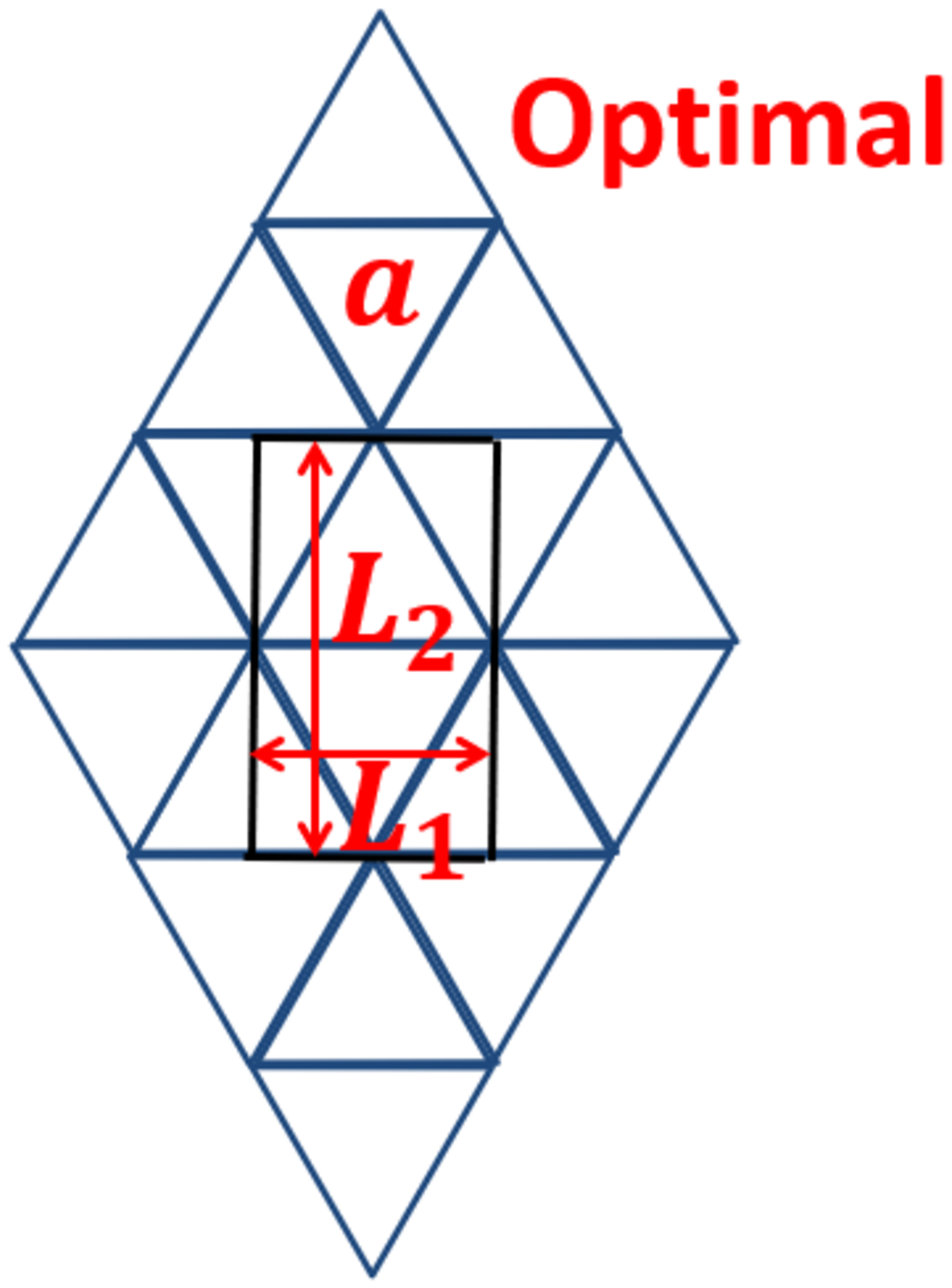} &
\includegraphics[width=0.25\textwidth]{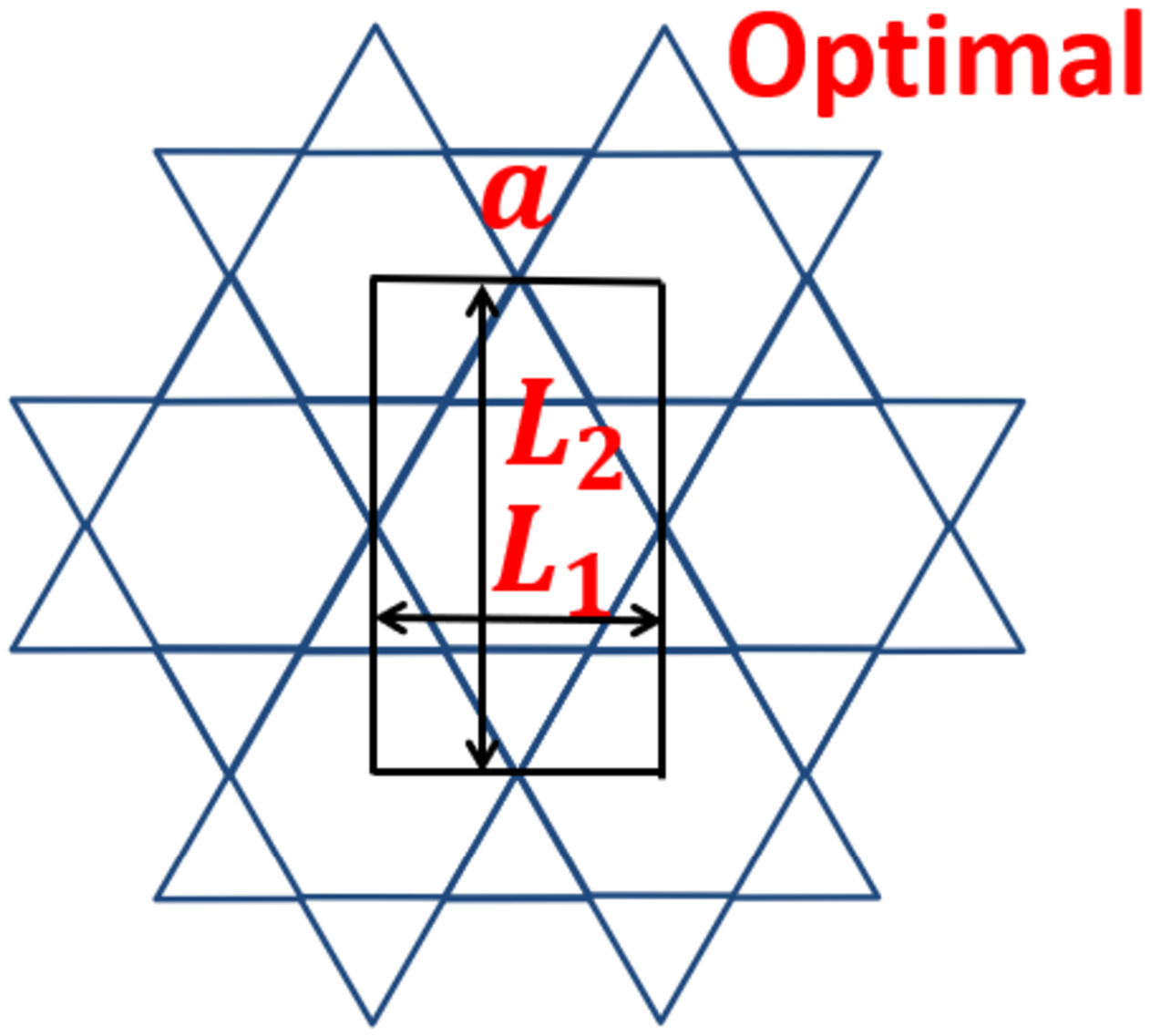} \\
\mbox{\bf (a)} & \mbox{\bf (b)} & \mbox{\bf (c)} \\
\includegraphics[width=0.25\textwidth]{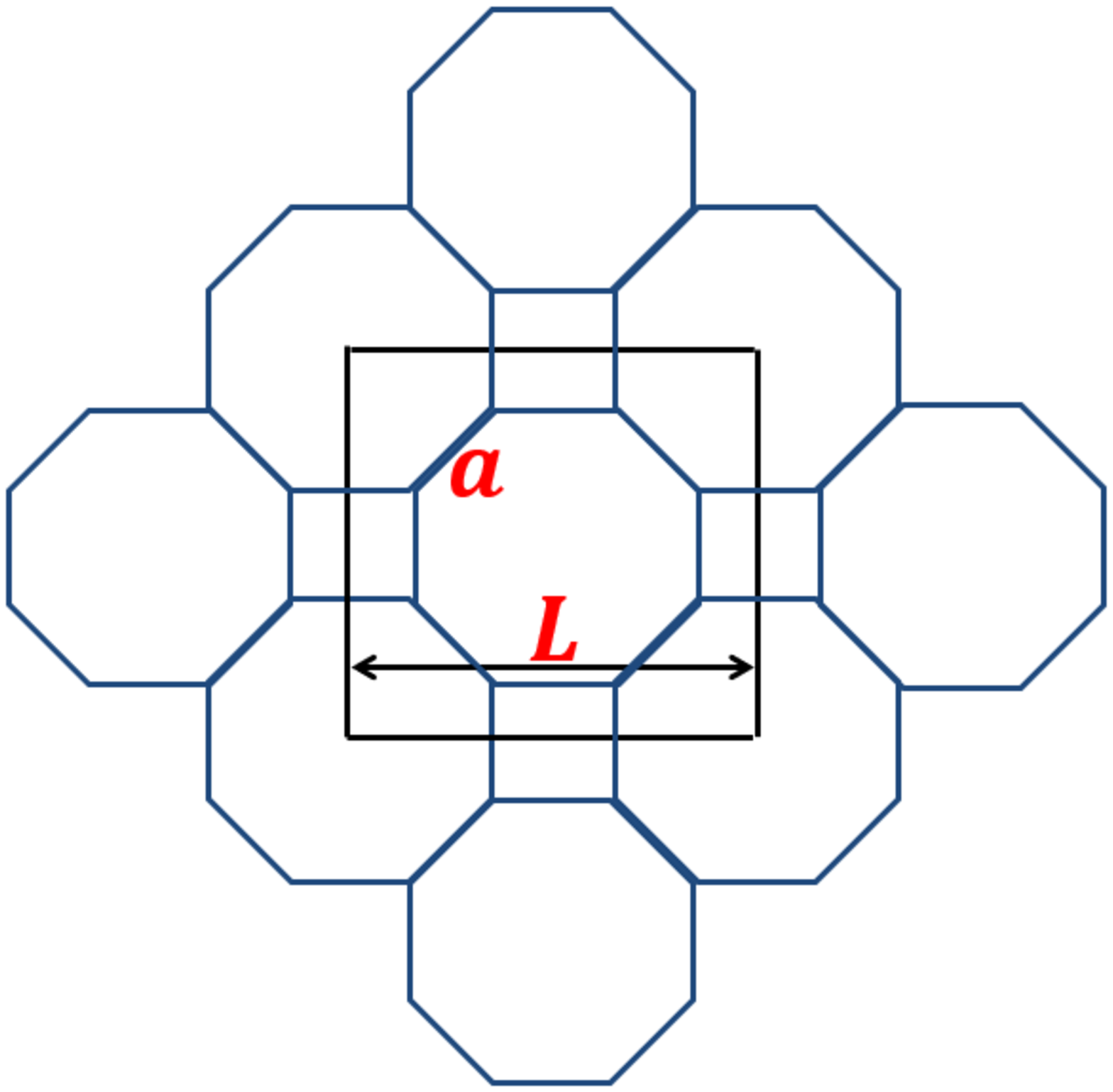} &
\includegraphics[width=0.25\textwidth]{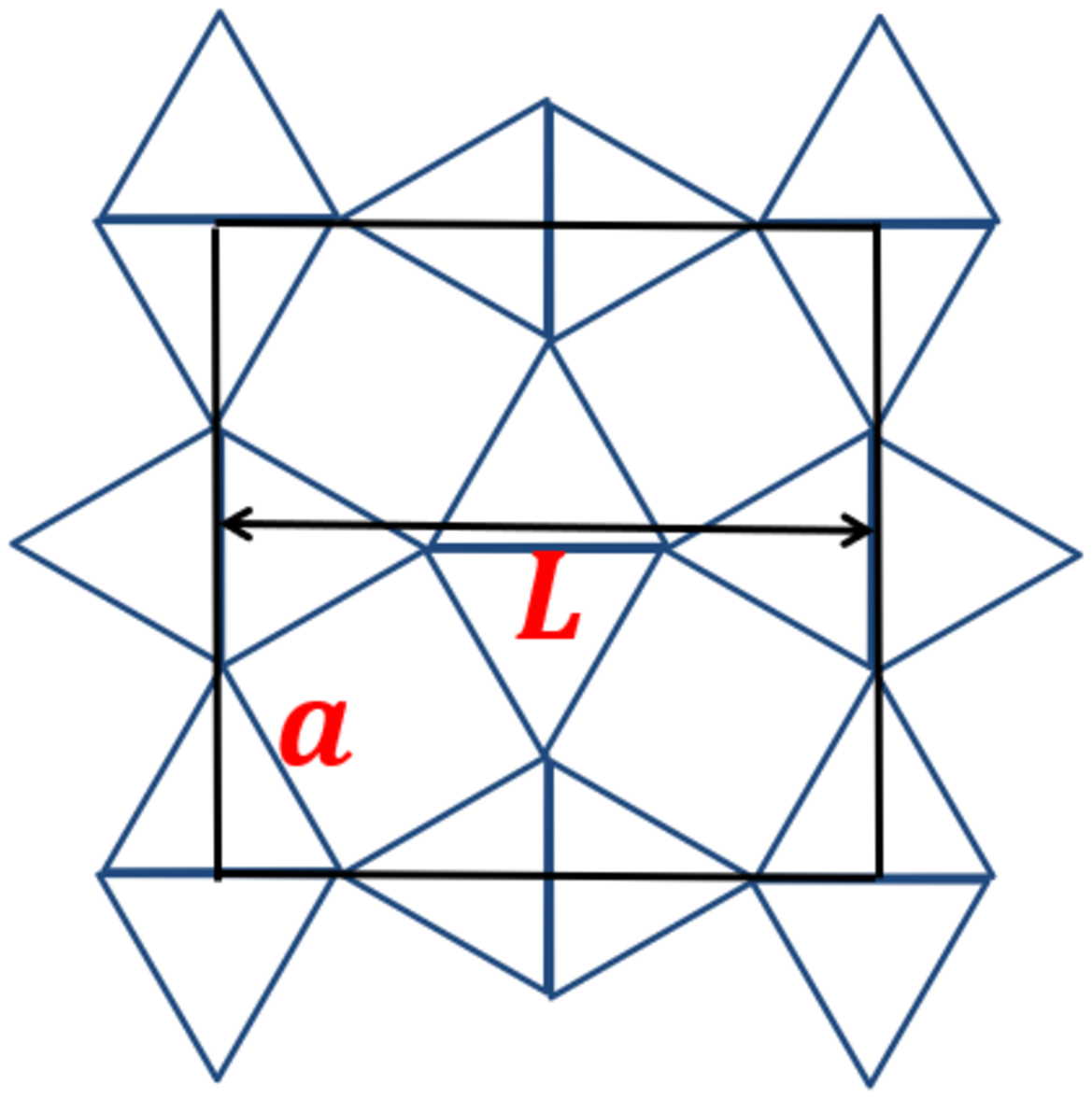} &
\includegraphics[width=0.25\textwidth]{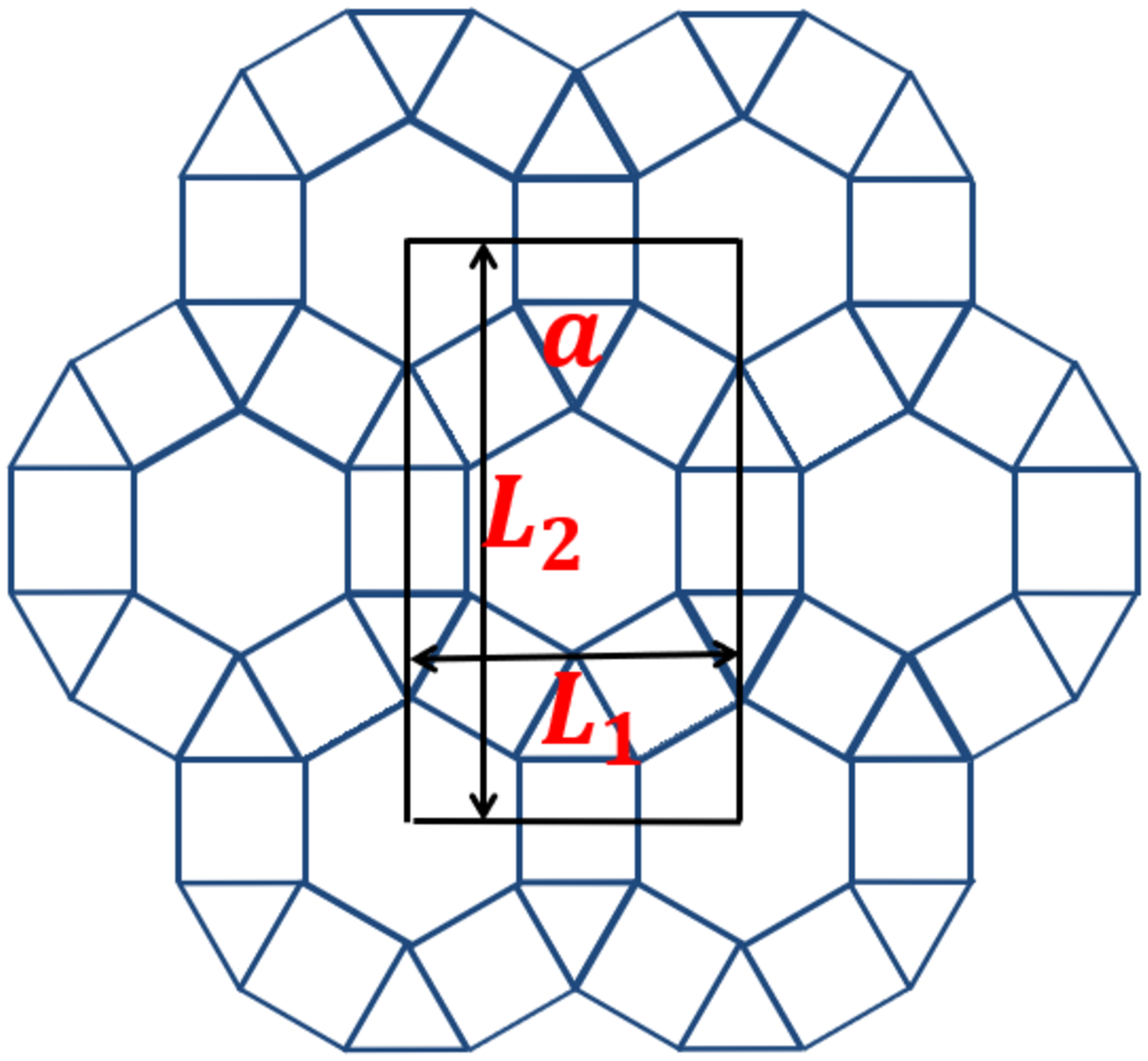} \\
\mbox{\bf (d)} & \mbox{\bf (e)} & \mbox{\bf (f)} \\
\includegraphics[width=0.25\textwidth]{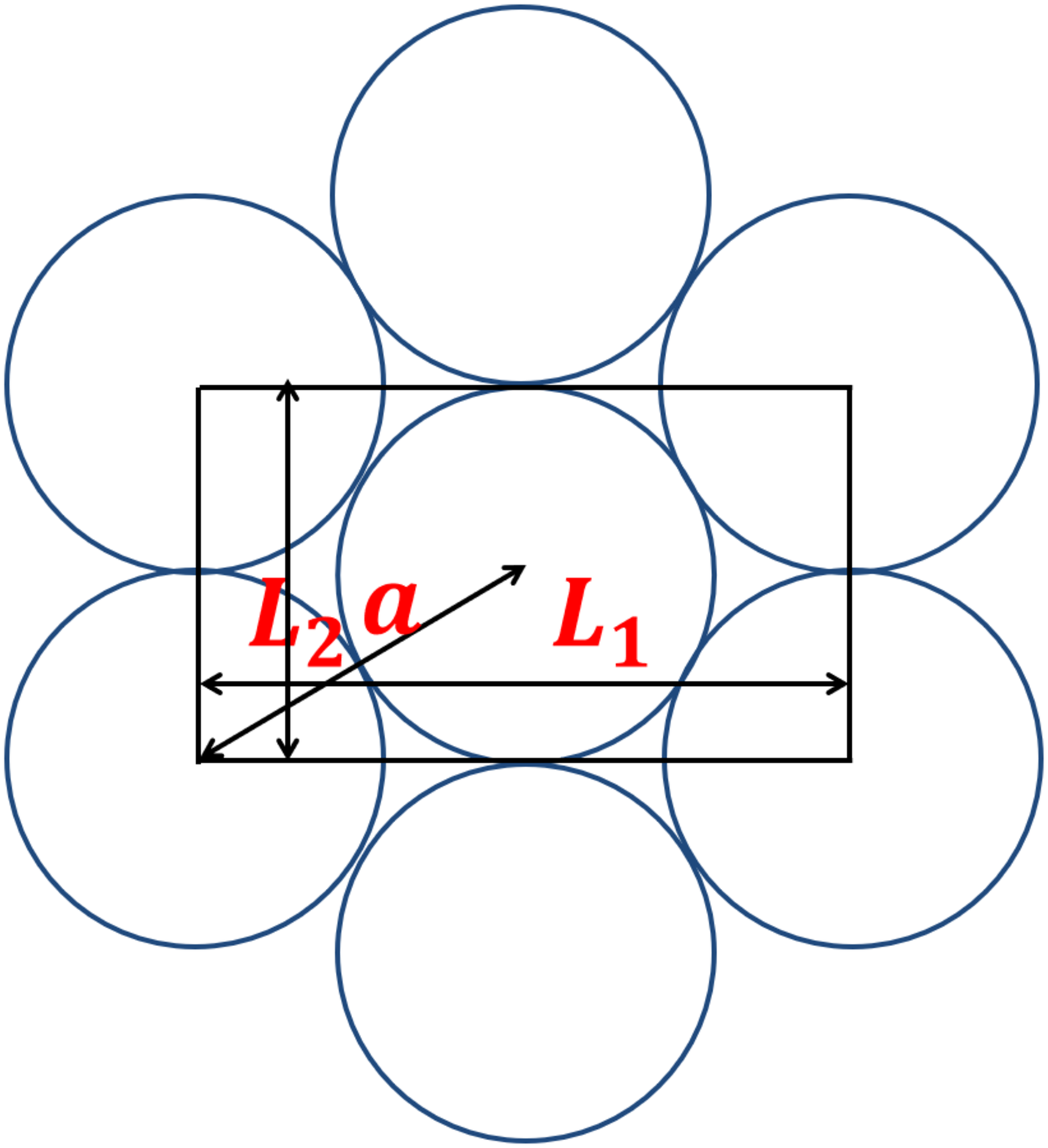} &
\includegraphics[width=0.30\textwidth]{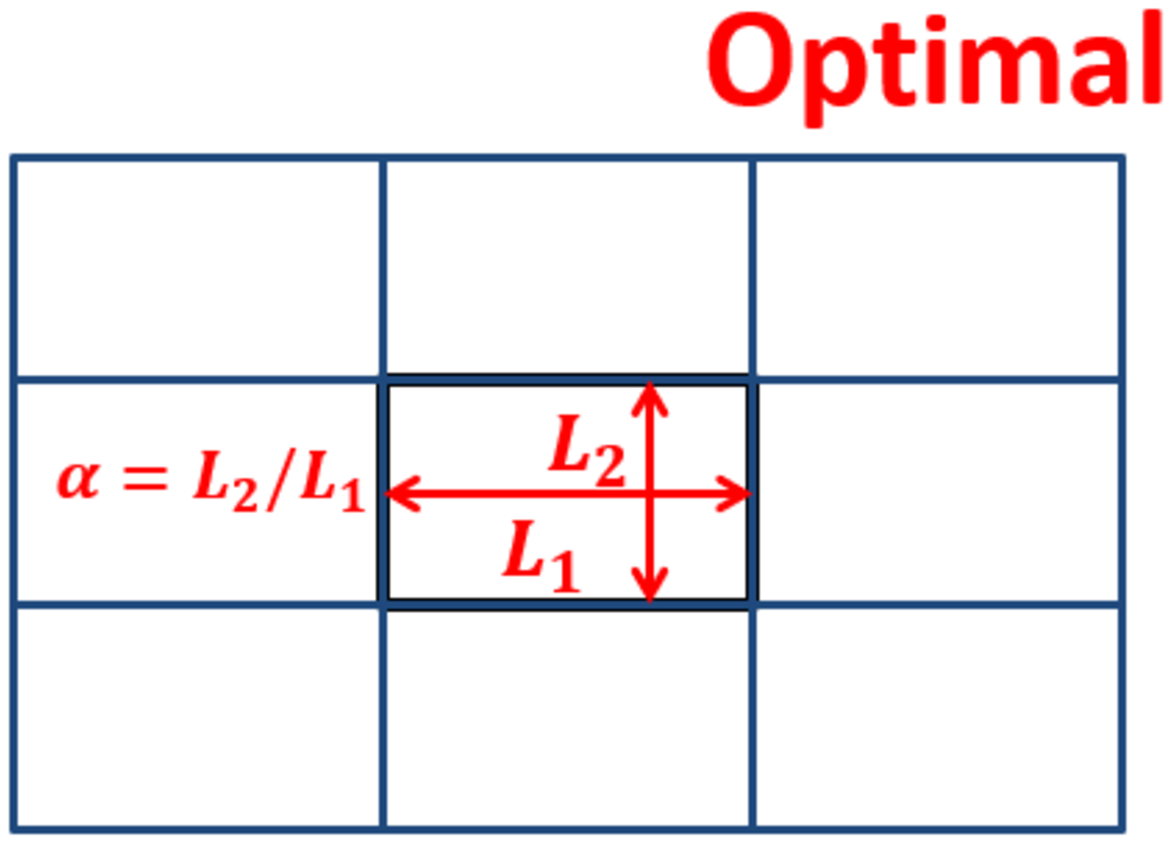} &
\includegraphics[width=0.18\textwidth]{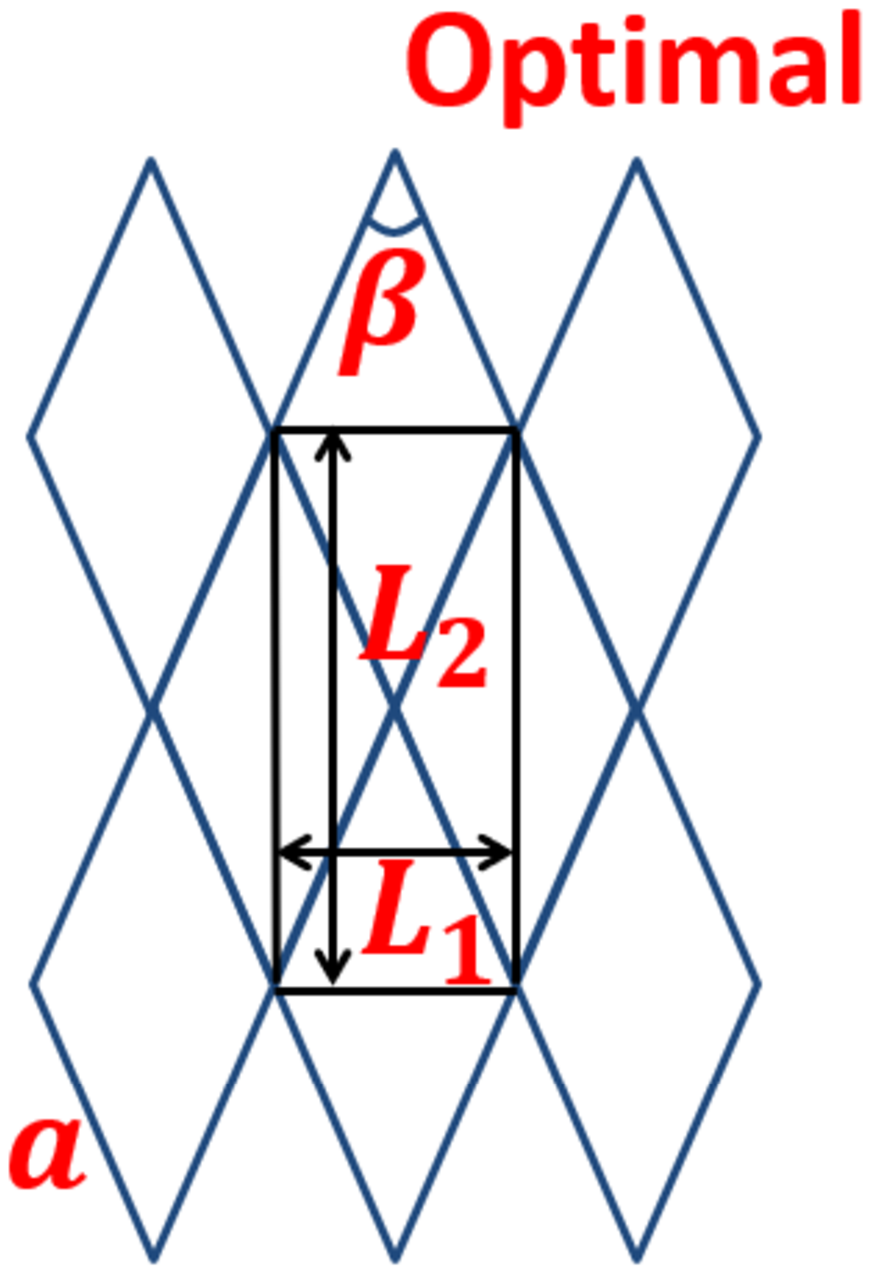} \\
\mbox{\bf (g)} & \mbox{\bf (h)} & \mbox{\bf (i)} \\
\end{array}$
\end{center}
\caption{Illustrations of various periodic cellular network structures and their fundamental cells. 
The fundamental cells are indicated in black. The lengths of the
fundamental cell in the $x_1$ and $x_2$ directions are denoted by $L_1$ and $L_2$, respectively. In certain cases where
$L_1 = L_2$, we simply use $L$ to denote both lengths. (a) Honeycomb network. (b) Triangular network.
(c) Kagom\'{e} network. (d) Octagonal network. (e) Snub square network. (f) Overlapping dodecagonal network. (g) Triangular lattice of circles.
(h) Rectangular network. Note that square network is a special case of 
rectangular network, where the aspect ratio $\alpha=1$. (i) Rhombic network. Note 
that square network is a special case of rhombic network, where $L_1=L_2$ and $\beta=\pi/2$.}
\label{fig_3}
\end{figure*}

\setlength\extrarowheight{5pt}
\begin{table*}[ht!]
\caption{Effective conductivity tensor $\rttensor{\sigma}_e$ and tortuosity tensor $\rttensor{\tau}$ of various periodic network structures as shown in Fig.
\ref{fig_3}. Note that the honeycomb, triangular, kagom\'{e}, square, rectangular and rhombic networks 
possess the optimal values of the effective conductivity.}
\begin{center}
\begin{tabular}{{c}{c}{c}} \\ \hline\hline
Network & $\rttensor{\sigma}_e/(\phi_2\sigma)$ & $\rttensor{\tau}$ \\
\hline
Honeycomb & $\begin{bmatrix} \frac{1}{2} & 0 \\ 0 & \frac{1}{2} \end{bmatrix}$ & $\begin{bmatrix} 1 & 0 \\ 0 & 1 \end{bmatrix}$ \\
\hline
Triangular & $\begin{bmatrix} \frac{1}{2} & 0 \\ 0 & \frac{1}{2} \end{bmatrix}$ & $\begin{bmatrix} 1 & 0 \\ 0 & 1 \end{bmatrix}$ \\
\hline
Kagom\'{e} & $\begin{bmatrix} \frac{1}{2} & 0 \\ 0 & \frac{1}{2} \end{bmatrix}$ & $\begin{bmatrix} 1 & 0 \\ 0 & 1 \end{bmatrix}$ \\
\hline
Octagonal & $\begin{bmatrix} \frac{3+2\sqrt{2}}{12} & 0 \\ 0 & \frac{3+2\sqrt{2}}{12} \end{bmatrix}$ & $\begin{bmatrix} 1.0294 & 0 \\ 0 & 1.0294 \end{bmatrix}$ \\
\hline
Snub square & $\begin{bmatrix} \frac{4+2\sqrt{3}}{15} & 0 \\ 0 & \frac{4+2\sqrt{3}}{15} \end{bmatrix}$ & $\begin{bmatrix} 1.0048 & 0 \\ 0 & 1.0048 \end{bmatrix}$ \\
\hline
Overlapping dodecagonal & $\begin{bmatrix} \frac{2+\sqrt{3}}{8} & 0 \\ 0 & \frac{2+\sqrt{3}}{8} \end{bmatrix}$ & $\begin{bmatrix} 1.0718 & 0 \\ 0 & 1.0718 \end{bmatrix}$ 
\\
\hline
Triangular lattice of circles & $\begin{bmatrix} \frac{9}{2\pi^{2}} & 0 \\ 0 & \frac{9}{2\pi^{2}} \end{bmatrix}$ & $\begin{bmatrix} 1.0966 & 0 \\ 0 & 1.0966 \end{bmatrix}$  \\
\hline
Rectangular & $\begin{bmatrix} \frac{1}{1+\alpha} & 0 \\ 0 & \frac{\alpha}{1+\alpha} \end{bmatrix}$ & $\begin{bmatrix} 1 & 0 \\ 0 & 1 \end{bmatrix}$ \\
\hline
Rhombic & $\begin{bmatrix} \sin^2(\frac{\beta}{2}) & 0 \\ 0 & \cos^2(\frac{\beta}{2}) \end{bmatrix}$ & $\begin{bmatrix} 1 & 0 \\ 0 & 1 \end{bmatrix}$ \\
\hline\hline
\end{tabular}
\end{center}
\label{table_1}
\end{table*}

To quantify how much the effective conductivity $\rttensor{\sigma}_e$ of a certain structure deviates 
from the upper bound $\rttensor{\sigma_{U}}^{(2)}$, or how ``tortuous'' the conduction path is, 
we introduce what we call the ``tortuosity'' tensor\footnote{Traditionally 
tortuosity has been defined to be a purely geometric scalar quantity: 
the ratio of the average length of the fluid paths and the
geometrical length of the sample \cite{Ma08}. Our new tortuosity tensor is distinguished
from earlier definition in that
it is based on the transport behavior (not purely geometrical features)
and anisotropic media.} $\rttensor{\tau}$: 
\begin{equation}
\label{eq_n4} 
\rttensor{\tau}=\begin{bmatrix} \tau_1 & 0 \\ 
0 & \tau_2 \end{bmatrix}
\end{equation}
Here $\tau_I (I = 1, 2)$ denotes the $I$-th eigenvalue of $\rttensor{\tau}$ and is given by
\begin{equation}
\label{eq_n5} \tau_I = (\rttensor{\sigma_{U}}^{(2)})_{II}/(\rttensor{\sigma}_e)_{II},
\end{equation} 
where $(\rttensor{\sigma_{U}}^{(2)})_{II}$ and $(\rttensor{\sigma}_e)_{II}$ are the $I$-th eigenvalues 
of $\rttensor{\sigma_{U}}^{(2)}$ and $\rttensor{\sigma}_e$, respectively. For macroscopically isotropic structures, 
the tortuosity reduces to a scalar quantity $\tau$.
Note that for optimal structures, the eigenvalues $\tau_1 = \tau_2 = 1$.

Using these procedures, we first determine the effective conductivities 
of ordered (periodic) hyperuniform networks shown in Fig. \ref{fig_3}, 
which include both macroscopically isotropic and anisotropic varieties.  
The computed effective conductivity and tortuosity tensors of these structures 
are listed in Table \ref{table_1}. Note that among all of the macroscopically isotropic structures investigated, 
the honeycomb network [Fig. \ref{fig_3}(a)], triangular network [Fig. \ref{fig_3}(b)], 
kagom\'{e} network [Fig. \ref{fig_3}(c)], and square network [a 
special case of the rectangular and rhombic networks in Fig. \ref{fig_3} (h) and (i)] possess the optimal value of the effective conductivity, 
i.e., they achieve the upper bound (\ref{eq_53}) \cite{Lu99}. The structures 
shown in Fig. \ref{fig_3} (d), (e), (f), and (g), on the other hand, possess 
suboptimal effective conductivities.
Note that the network consisting of
touching circles shown in Fig. \ref{fig_3} (g) 
possesses ``dead ends'', a structural feature that leads to a suboptimal effective conductivity. Indeed, 
this network possesses the lowest effective conductivity $\sigma_e$, or the highest scalar 
tortuosity $\tau$, among all of the networks investigated in this study. 

\subsection{Effective Conductivity of Intersecting Parallel-Channel Cellular Structures}
\begin{figure}[ht!]
\begin{center}
$\begin{array}{c}
\includegraphics[width=0.35\textwidth]{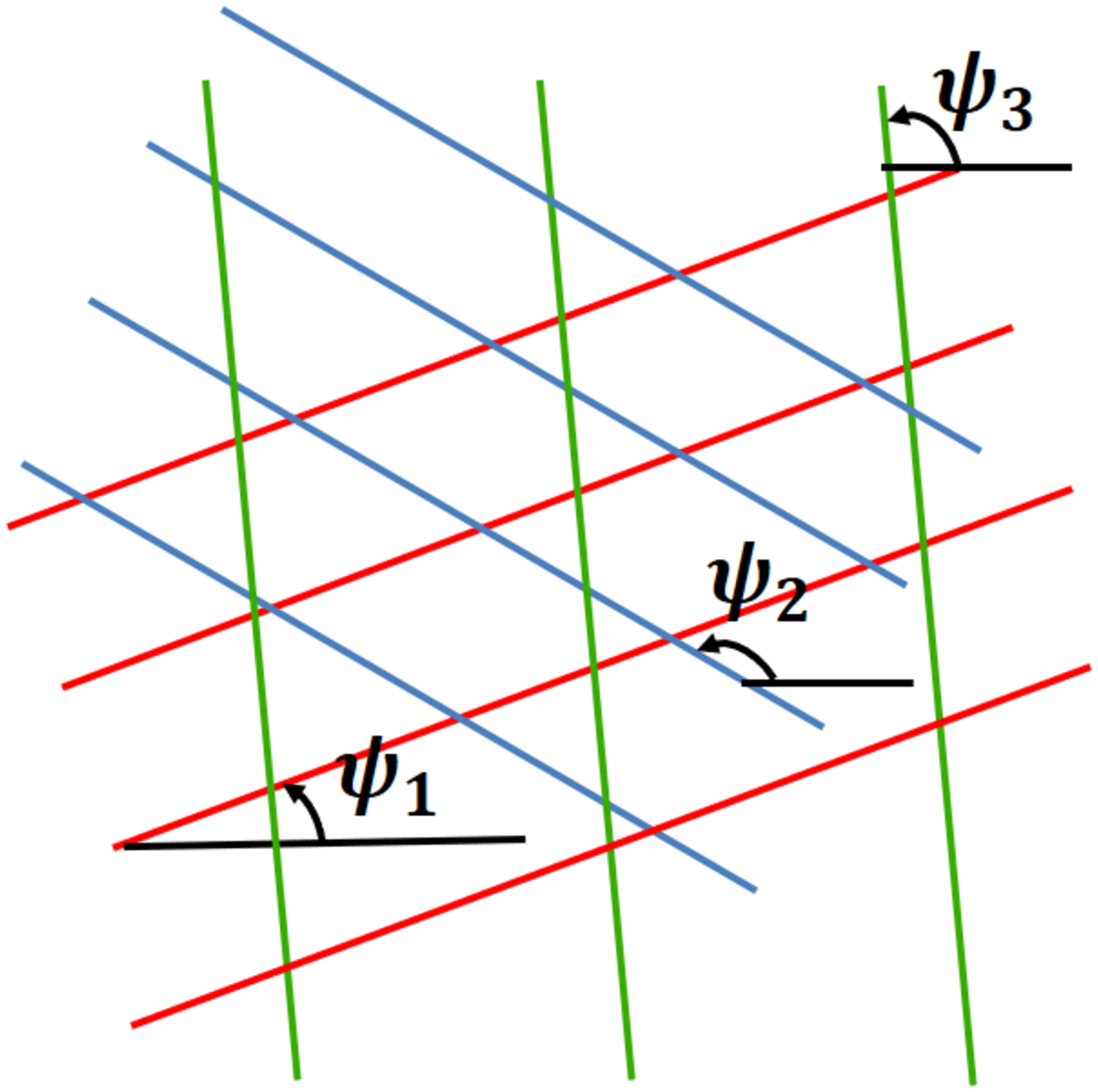} \\
\mbox{\bf (a)} \\
\includegraphics[width=0.35\textwidth]{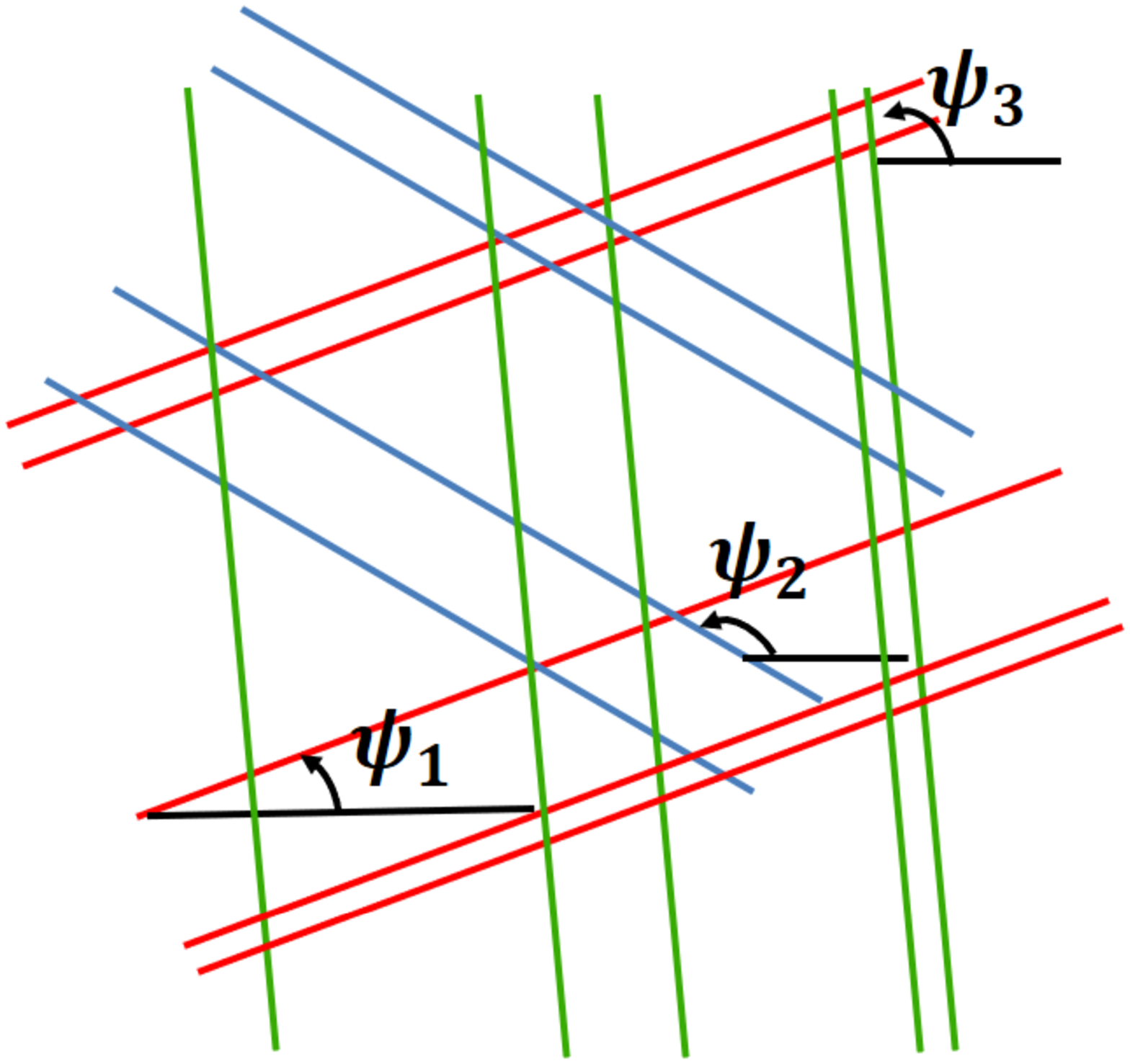} \\
\mbox{\bf (b)} 
\end{array}$
\end{center}
\caption{Schematics of ordered (a) and disordered (b)
cellular structures consisting of three sets of 
intersecting parallel channels oriented in directions 
with polar angles $\psi_1, \psi_2$ and
$\psi_3$, respectively. We stress that 
the parallel channels in each set are not required to be 
equally spaced, and thus the networks discussed here 
include disordered varieties.}
\label{fig_4}
\end{figure}

We now consider
cellular structures that are constructed by superposing $N (N\geq2)$ sets of intersecting parallel channels oriented in directions with polar angles $\psi_1, \psi_2, ..., \psi_N$, respectively, as
schematically shown in Fig. \ref{fig_4}. We stress that 
the parallel channels in each set are not required to be 
equally spaced, and thus the networks discussed here 
include disordered ones [see Fig. \ref{fig_4}(b) for an example]. Note that the rectangular and rhombic networks shown in Figs. \ref{fig_3} (h) and (i) are special 
examples of this type of structures.
The relative volume fraction of the $i$-th set of channels is denoted by $c_i$ ($i=1,2,..., N$, and $\sum\limits_{i=1}^{N}c_i=1$), where $c_i = \phi_{2,i}/\phi$ and $\phi_{2, i}$ is the volume fraction of the $i$-th set of channels.

For such an intersecting parallel-channel network, 
we can compute the effective conductivity $\T{\sigma}_e$ exactly. 
Specifically, application of the procedure described in Sec. 5.1 
to such a general structure yields the following 
effective conductivity $\T{\sigma}_e$:
\begin{equation}
\label{eq_n2} \T{\sigma}_e=\begin{bmatrix} \sum\limits_{i=1}^{N}\cos^2(\psi_i)c_i &
\sum\limits_{i=1}^{N}\cos(\psi_i)\sin(\psi_i)c_i \\ \sum\limits_{i=1}^{N}\cos(\psi_i)\sin(\psi_i)c_i &
\sum\limits_{i=1}^{N}\sin^2(\psi_i)c_i \end{bmatrix}\phi.
\end{equation}

\subsection{Demonstration of Optimality for Intersecting Parallel-Channel Cellular Structures}
We now prove that the effective conductivity tensor (\ref{eq_n2}) for
any intersecting parallel-channel network is optimal by showing that
it corresponds to the the upper bound $\rttensor{\sigma_{U}}^{(2)}$ on $\T{\sigma}_e$.
We begin by computing the two-point tensor coefficient $\mathbf{A}$ of such a network, 
which is explicitly given by
\begin{equation}
\label{eq_54} \mathbf{A}=\frac{1}{\pi}\lim_{\delta \rightarrow 0}
\int_{\delta}^{\infty}\frac{dr}{r}\int_{0}^{2\pi}d\theta\chi(r, \theta)\begin{bmatrix} \cos(2\theta) & \sin(2\theta) \\ \sin(2\theta)
& -\cos(2\theta) \end{bmatrix}
\end{equation}
where $\chi(r, \theta) = S_{2}(r,\theta)-\phi^2$ is the autocovariance function of the cellular network, and
$S_{2}(r,\theta)$ is the two-point correlation function of the channel phase. The autocovariance function
$\chi(r, \theta)$ can be decomposed into two parts:
\begin{equation}
\label{eq_55} \chi(r, \theta) = \sum\limits_{i=1}^{N}[S_{2, ii}(r, \theta) - \phi_{2, i}^{2}] +
\sum\limits_{i<j}[S_{2, ij}(r, \theta) - 2\phi_{2, i}\phi_{2, j}],
\end{equation}
where the first part corresponds to the two-point correlation between channels in the same set, and the second part
corresponds to the two-point cross-correlation between channels in different sets, and $\phi_{2, i}$ is the
volume fraction the $i$-th set of channels. Since the two-point cross-correlation term $S_{2, ij}(r, \theta) -
2\phi_{2, i}\phi_{2, j}$ depends only on the distance $r$, i.e., independent of the orientation $\theta$, the
contribution to $\mathbf{A}$ from the second part is $\mathbf{0}$. By 
adding up the contributions to $\mathbf{A}$ from the self-correlation of each individual set of intersecting parallel channels,
we find
\begin{equation}
\label{eq_56} \mathbf{A}=\begin{bmatrix} \sum\limits_{i=1}^{N}\cos(2\psi_i)c_i &
\sum\limits_{i=1}^{N}\sin(2\psi_i)c_i \\ \sum\limits_{i=1}^{N}\sin(2\psi_i)c_i &
-\sum\limits_{i=1}^{N}\cos(2\psi_i)c_i \end{bmatrix}\phi.
\end{equation}
We can diagonalize the above matrix to obtain the eigenvalues for $\mathbf{A}$ once the relative volume fractions
and orientations of each set of intersecting parallel channels are given. In general, the ``superposition'' of sets of intersecting parallel channels
produces a macroscopically anisotropic structure, and the corresponding $\mathbf{A}$ is not $\mathbf{0}$.

By substituting $\mathbf{A}$, given by Eq. (\ref{eq_56}), into Eq. (\ref{eq_36}), we 
see that the upper bound $\rttensor{\sigma_{U}}^{(2)}$ for these structures is exactly 
the same as $\T{\sigma}_e$ given by Eq. (\ref{eq_n2}). Thus, 
we have rigorously demonstrated that anisotropic structures consisting of sets of intersecting
parallel channels achieve the two-point anisotropic generalizations 
of the Hashin--Shtrikman bound (\ref{eq_19})
on $\T{\sigma}_e$, regardless of whether they are 
ordered or disordered, hyperuniform or nonhyperuniform.

In addition, we note that in certain special cases, where the $N$ sets of intersecting parallel channels have identical relative volume fraction, i.e.,
$c_i=\frac{1}{N}(i=1,2,...,N)$, and the channels are superpositioned in a way such that the overall structure possesses
$N$-fold rotational symmetry, we can show that $\mathbf{A}=\mathbf{0}$. Specifically, without loss of generality, we
can have one set of channels aligned with the horizontal axis, and the other sets oriented in directions with polar angles
$\psi = \frac{\pi}{N}, ..., \frac{\pi(N-1)}{N}$, respectively, with $\mathbf{A}$ now given by
\begin{equation}
\label{eq_57} \mathbf{A}=\begin{bmatrix} \sum\limits_{i=0}^{N-1}\cos(\frac{2\pi}{N}i) &
\sum\limits_{i=0}^{N-1}\sin(\frac{2\pi}{N}i) \\ \sum\limits_{i=0}^{N-1}\sin(\frac{2\pi}{N}i) &
-\sum\limits_{i=0}^{N-1}\cos(\frac{2\pi}{N}i) \end{bmatrix}\frac{\phi}{N}=\mathbf{0}.
\end{equation}
Thus, the resulting structure is macroscopically isotropic.

\subsection{Cross-Property Relations}
For periodic cellular structures with 3-, 4- or 6-fold rotational symmetry, 
the cross-property bound (\ref{eq_40}) allows us to obtain upper
bounds on the effective bulk moduli given the measurement of the effective conductivity
of the structures. Interestingly, whenever the effective conductivity $\sigma_e$ of 
certain structure is optimal, so are the effective bulk moduli. The results are summarized in Table \ref{table_2}. 
Note that the square, honeycomb, and kagom\'{e} 
networks possess optimal effective bulk moduli, i.e., they achieve
the upper bound (\ref{eq_40}).

\begin{table}[ht!]
\caption{Upper bounds on the effective moduli $K_e$ of certain periodic network structures in Fig. \ref{fig_3}, which are scaled by modulus $K$, and volume fraction $\phi$ of the ``channel'' phase. It is noteworthy that the square, honeycomb and kagom\'{e} 
networks possess optimal $K_e$.}
\begin{center}
\begin{tabular}{{c}{c}{c}{c}} \\ \hline\hline
Network & $K_{e}/(K\phi)$  \\
\hline
Honeycomb & $0.5(1-\nu)$ \\
\hline
Triangular & $0.5(1-\nu)$ \\
\hline
Kagom\'{e} & $0.5(1-\nu)$ \\
\hline
Square & $0.5(1-\nu)$ \\
\hline
Octagonal & $\frac{3+2\sqrt{2}}{12}(1-\nu)$ \\
\hline
Snub square & $\frac{4+2\sqrt{3}}{15}(1-\nu)$ \\
\hline
Overlapping dodecagonal & $\frac{2+\sqrt{3}}{8}(1-\nu)$ \\
\hline
Triangular lattice of circles & $\frac{9}{2\pi^{2}}(1-\nu)$ \\
\hline\hline
\end{tabular}
\end{center}
\label{table_2}
\end{table}

\subsection{Results for Arbitrary Phase Contrast}
\begin{figure}[ht!]
\begin{center}
$\begin{array}{c}
\includegraphics[width=0.35\textwidth]{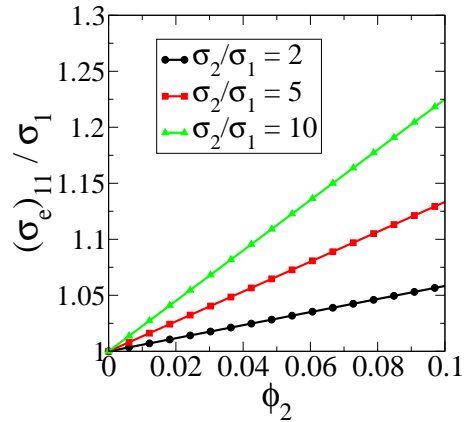} \\
\mbox{\bf (a)} \\
\includegraphics[width=0.35\textwidth]{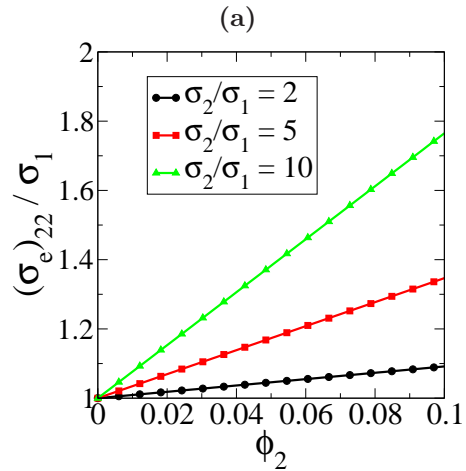} \\
\mbox{\bf (b)} 
\end{array}$
\end{center}
\caption{The principal components of the effective conductivity $({\sigma}_e)_{11}$ (a)
and $({\sigma}_e)_{22}$ (b) of oriented singly-coated space-filling ellipsoidal 
assemblages shown in Fig. \ref{fig_1} with an
aspect ratio $\alpha = 5.0$ as a function of the 
volume fraction of the more conducting phase $\phi_2$ 
at phase contrast ratios $\sigma_2/\sigma_1=2.0$, $5.0$, and $10.0$, as computed from 
Eq. (\ref{eq_35}). The effective conductivity of this anisotropic structure possesses optimal values.}  
\label{fig_7}
\end{figure}

It should not go unnoticed that many of the aforementioned
results are straightforwardly extended to cases in which
the void or matrix phase has non-zero phase properties.
In such instances, the lower bounds (\ref{eq_34}) no longer vanish.
Note that whenever the network structure
is optimal (i.e., maximizes the effective conductivity), the upper bound (\ref{eq_35}) on effective conductivity is an exact
result (i.e., achieved by certain structures) for arbitrary phase contrast.
For example, in Fig. \ref{fig_7}, we plot
the effective conductivity for the optimal case of the 
aforementioned oriented singly-coated space-filling ellipsoidal 
assemblages shown in Fig. \ref{fig_1} with an 
aspect ratio $\alpha = 5.0$ as a function of the 
volume fraction of the more conducting phase $\phi_2$ 
at phase contrast ratios $\sigma_2/\sigma_1=2.0$, $5.0$, and $10.0$.

\section{Effective Conductivity and Elastic Moduli of Hyperuniform and Nonhyperuniform Disordered Networks}

In this section, we determine the effective conductivity and 
elastic moduli of various statistically isotropic disordered 
hyperuniform and nonhyperuniform networks. Our goal is to investigate how hyperuniformity 
affects the effective conductivity and elastic moduli, 
and how close these effective properties of disordered hyperuniform networks can come to 
being optimal.

\subsection{Mapping Disordered Point Patterns to Disordered Networks}

We map various 2D disordered nonhyperuniform and hyperuniform point patterns into 2D cellular network structures by the three types of tessellations mentioned in Sec. 2.3: Delaunay, Voronoi and Delaunay-centroidal tessellations. 
We then compute the effective conductivities of the networks. These point patterns include
nonhyperuniform and hyperuniform ones in square domains subject to periodic boundary conditions.
For nonhyperuniform point patterns, we consider Poisson point patterns (which are uncorrelated on all length scales) and those associated with the centroids 
of equal-sized hard disks in packings generated by the random-sequential-addition (RSA) process \cite{Zh13} with $N=100$ points in each pattern. 
We consider hyperuniform point patterns associated with the centroids
of equal-sized hard disks in maximally-random-jammed (MRJ) packings \cite{At14} with $N=100$ points in each pattern, and
various disordered stealthy hyperuniform ones with different $\chi$ values \cite{Uc04,Zh15} and $N=150$ points in each pattern. These stealthy point patterns are generated using the procedure described in 
Ref. \cite{Zh15}. Specifically, an optimization objective function that targets the 
structure factor $S(k)$ to be exactly zero for a range of small wavenumbers is employed, 
which guarantees the stealthiness of the resulting point patterns.
As mentioned above, when $0<\chi<0.5$, the point pattern is disordered and henceforth
 we will employ point patterns with $\chi$ values in this range. Specifically, we pick three $\chi$ values: 0.3, 0.4, and 0.49 \cite{Uc04,Zh15}.

\begin{figure*}[ht!]
\begin{center}
$\begin{array}{c@{\hspace{0.1cm}}c@{\hspace{0.1cm}}c}\\
\includegraphics[width=0.32\textwidth]{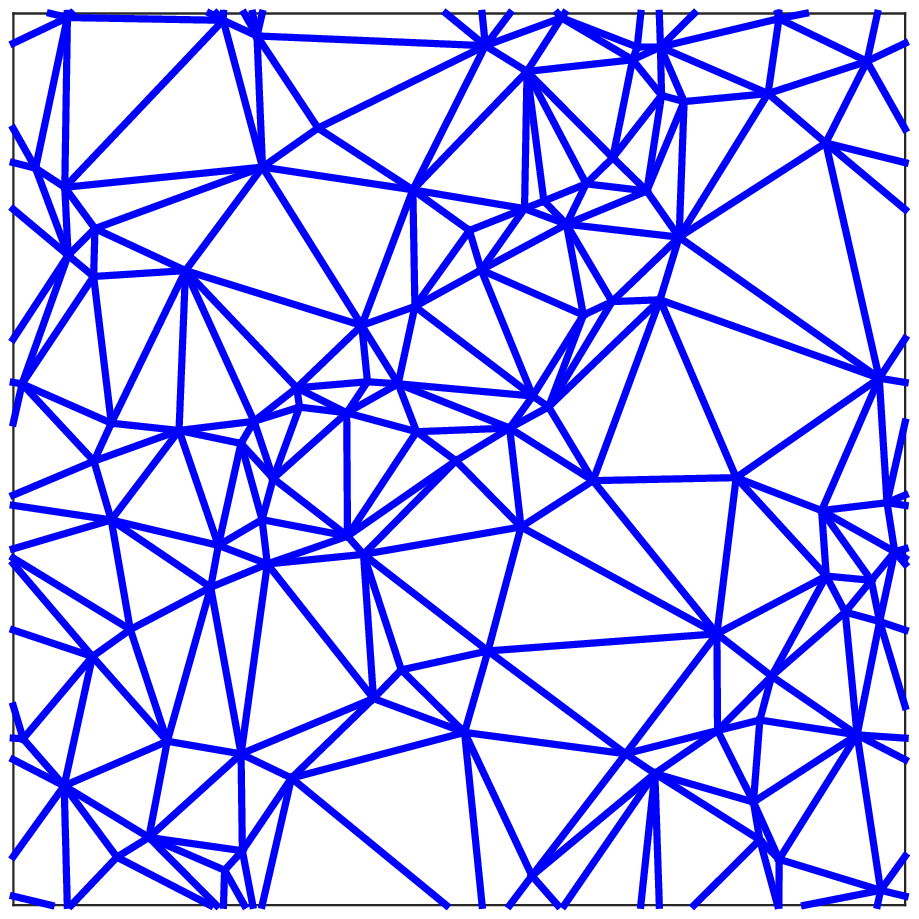} &
\includegraphics[width=0.32\textwidth]{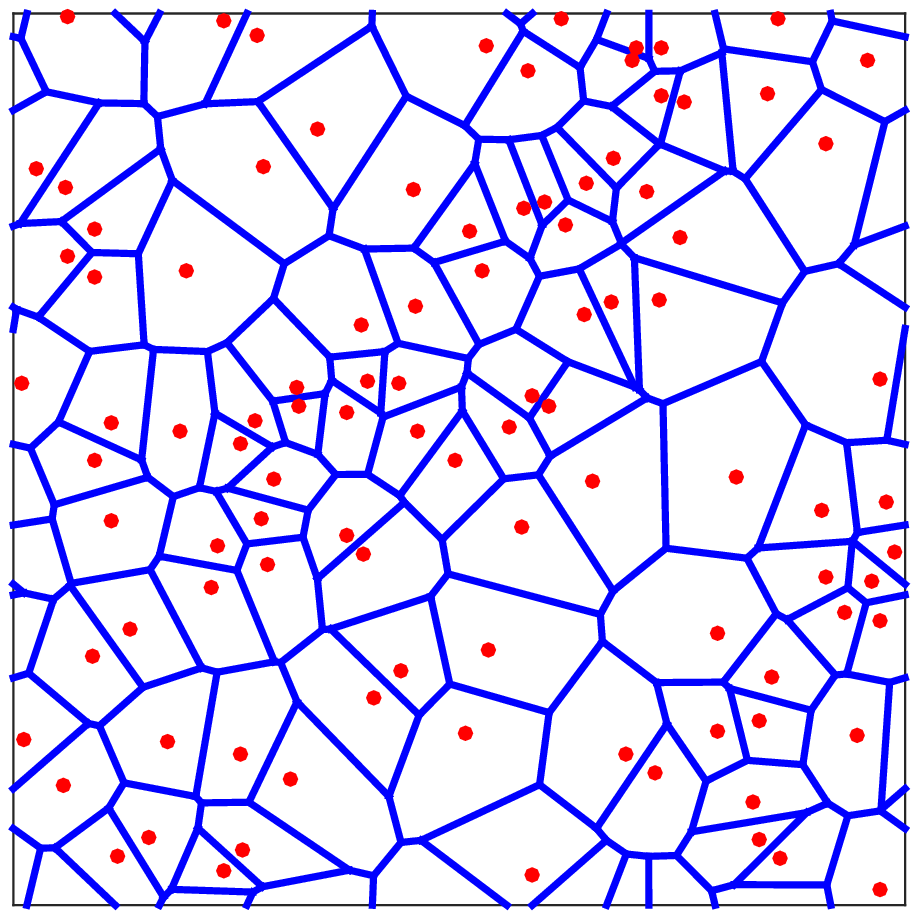} &
\includegraphics[width=0.32\textwidth]{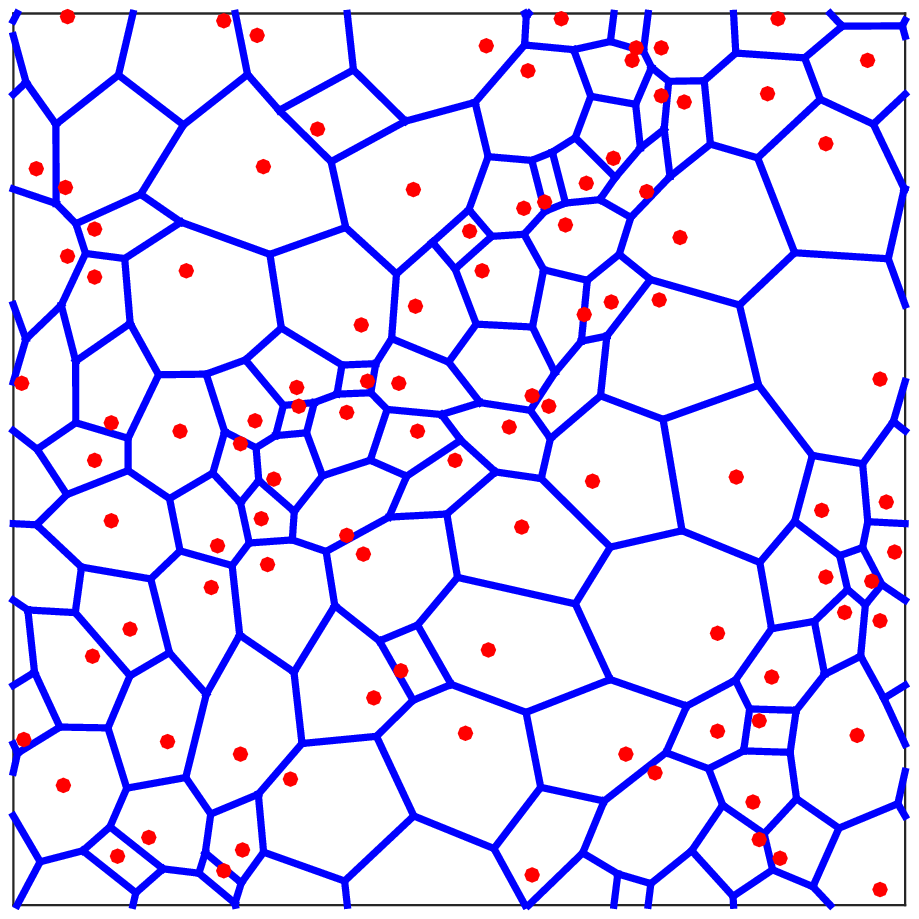} \\
\mbox{\bf (a)} & \mbox{\bf (b)} & \mbox{\bf (c)} \\
\includegraphics[width=0.32\textwidth]{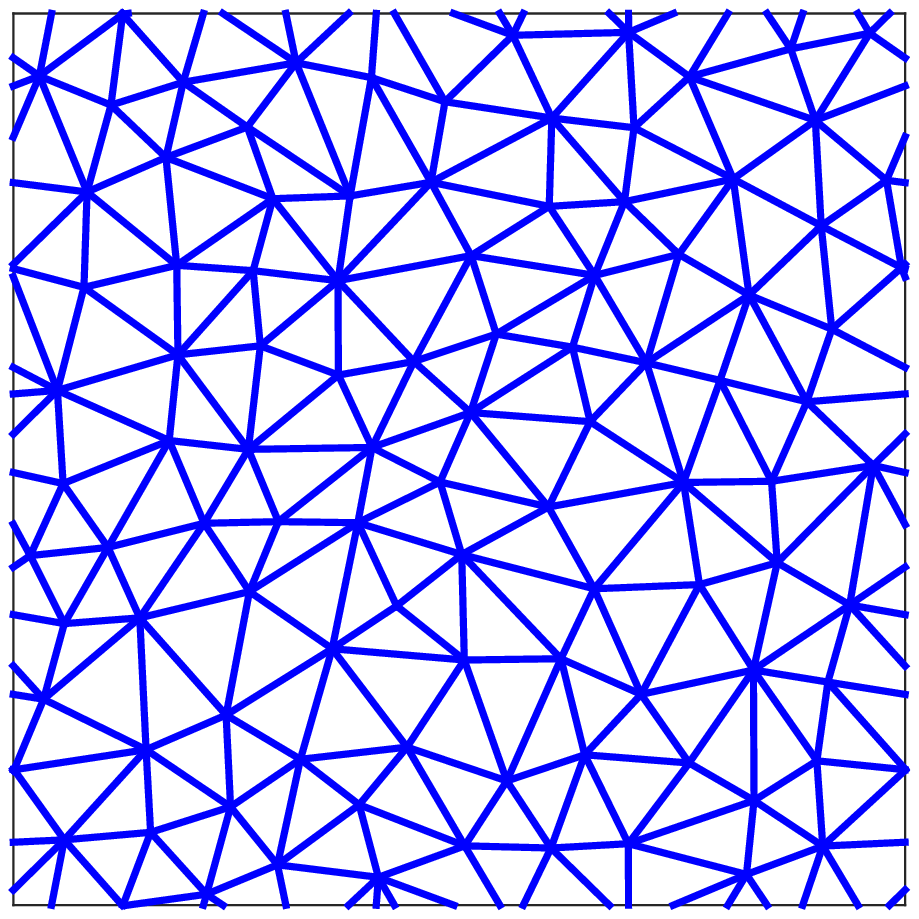} &
\includegraphics[width=0.32\textwidth]{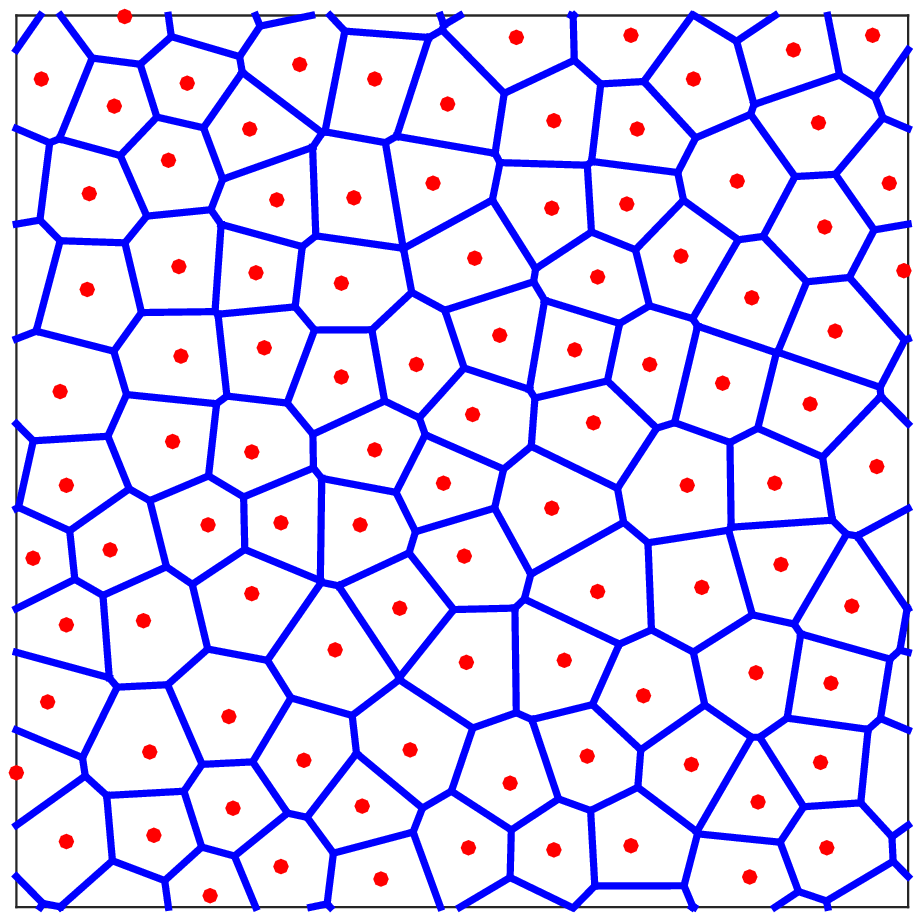} &
\includegraphics[width=0.32\textwidth]{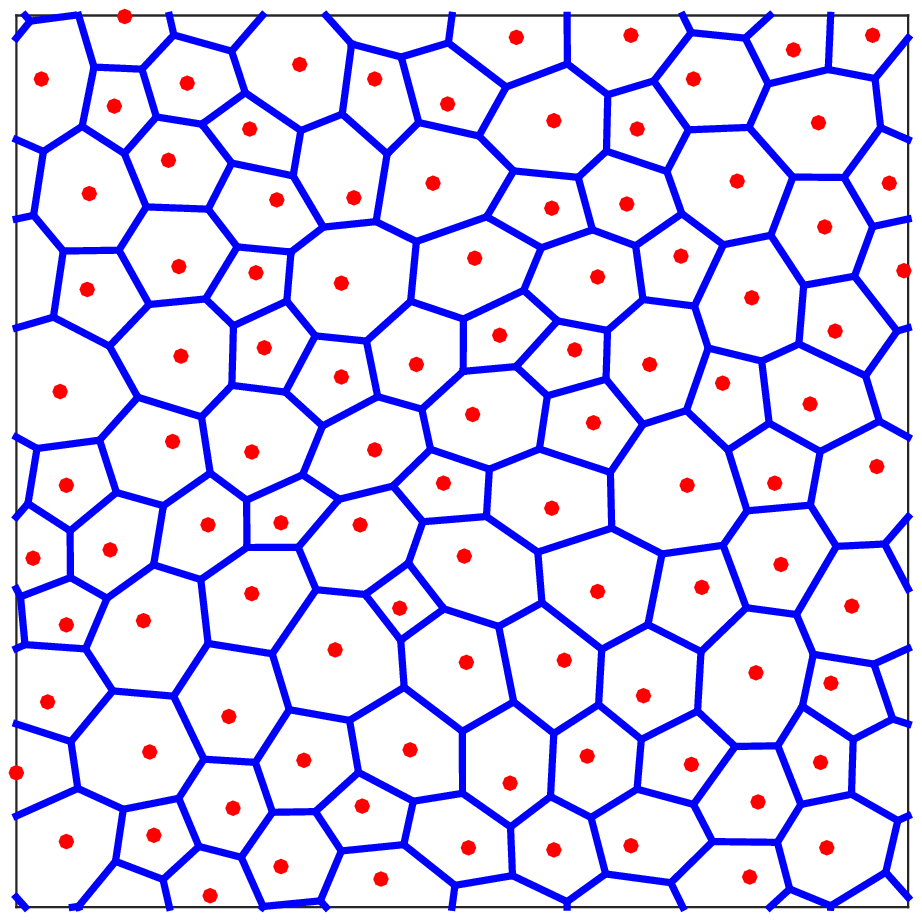} \\
\mbox{\bf (d)} & \mbox{\bf (e)} & \mbox{\bf (f)} \\
\includegraphics[width=0.32\textwidth]{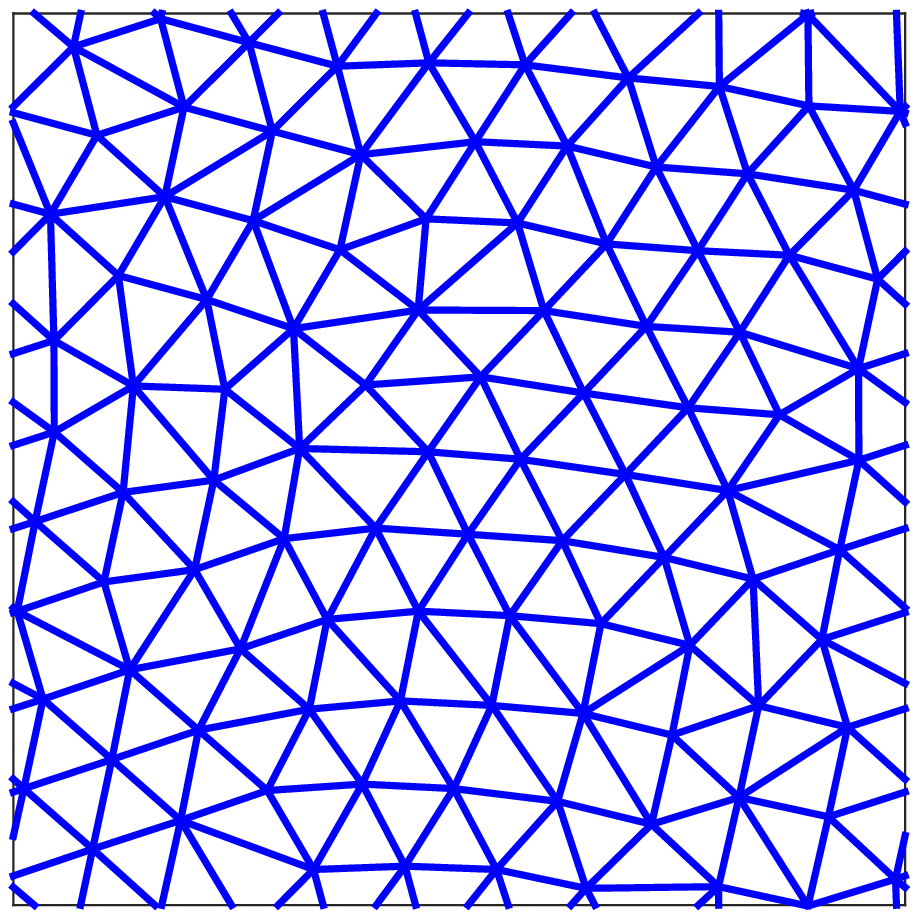} &
\includegraphics[width=0.32\textwidth]{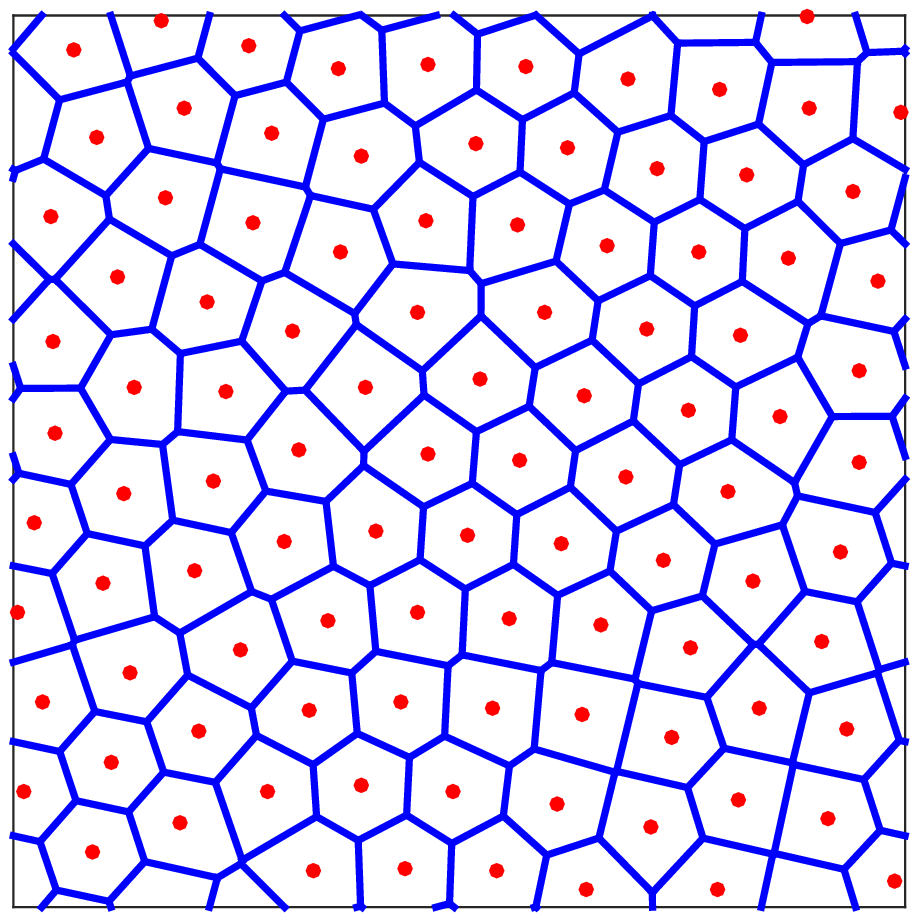} &
\includegraphics[width=0.32\textwidth]{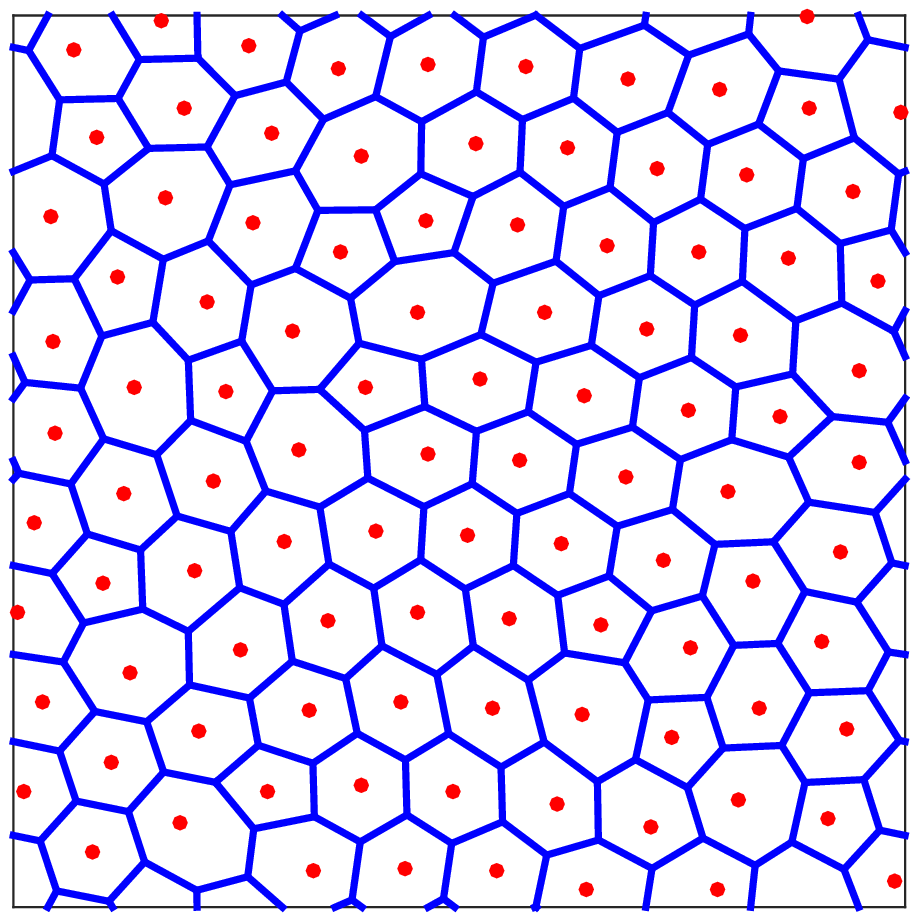} \\
\mbox{\bf (g)} & \mbox{\bf (h)} & \mbox{\bf (i)} \\
\end{array}$
\end{center}
\caption{Representative disordered nonstealthy cellular network structures mapped from various point patterns.
There are $N = 100$ points in each underlying point pattern. Note that in those Voronoi and Delaunay-centroidal networks,
the underlying point patterns are colored in red, and the conducting ``channels'' are colored in blue. In those Delaunay
networks, the points in the underlying point patterns are just the vertices of the triangles, which are colored in blue.
(a) Delaunay network of Poisson point pattern. (b) Voronoi network of Poisson point pattern. (c) Delaunay-centroidal network of
Poisson point pattern. (d) Delaunay network of RSA point pattern. (e) Voronoi network of RSA point pattern. (f) Delaunay-centroidal
network of RSA point pattern. (g) Delaunay network of MRJ point pattern. (h) Voronoi network of MRJ point pattern.
(i) Delaunay-centroidal network of MRJ point pattern.} \label{fig_5}
\end{figure*}

\begin{figure*}[ht!]
\begin{center}
$\begin{array}{c@{\hspace{0.1cm}}c@{\hspace{0.1cm}}c}\\
\includegraphics[width=0.32\textwidth]{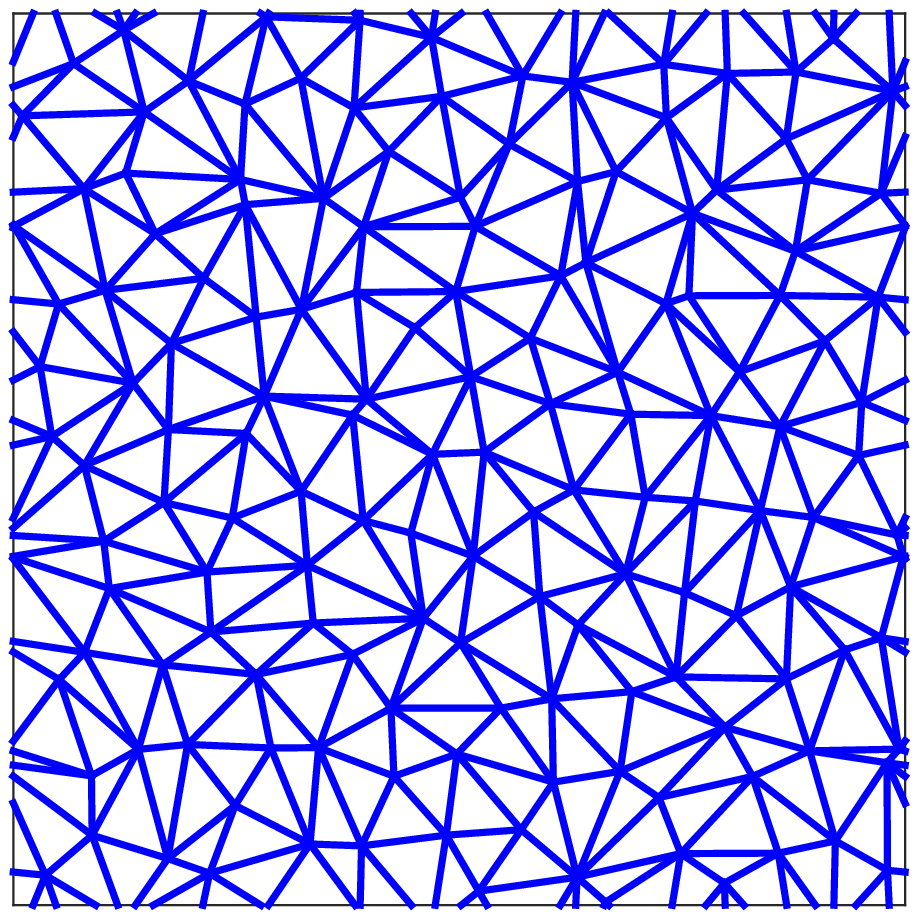} &
\includegraphics[width=0.32\textwidth]{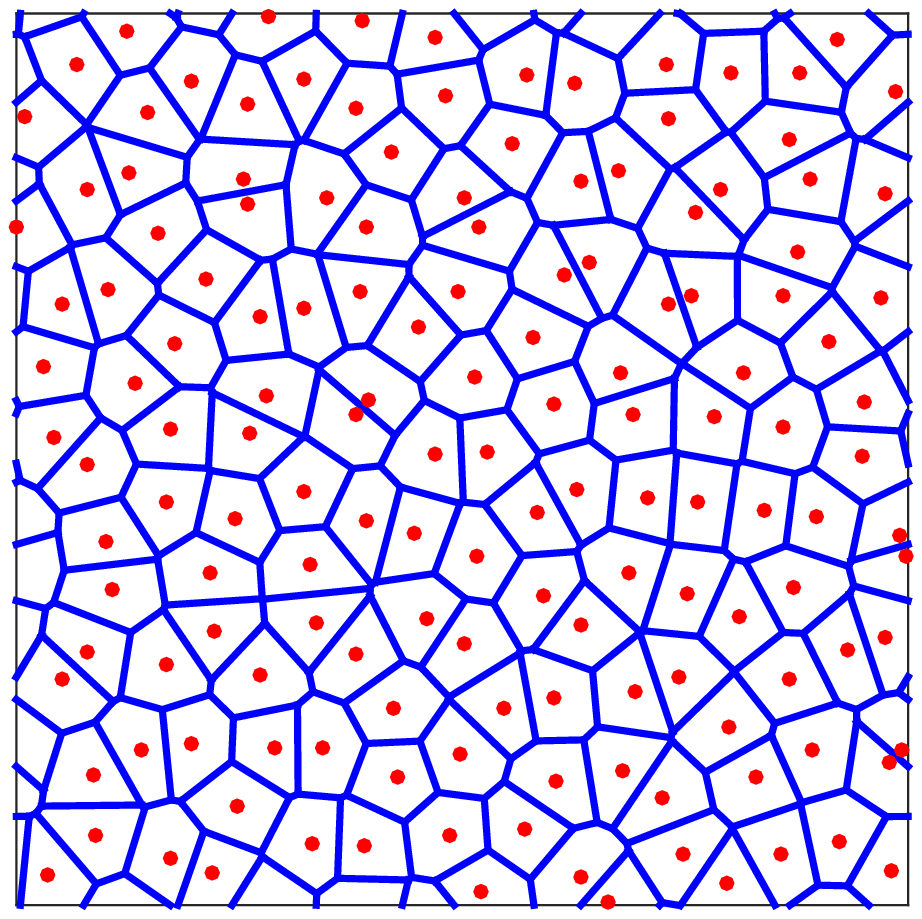} &
\includegraphics[width=0.32\textwidth]{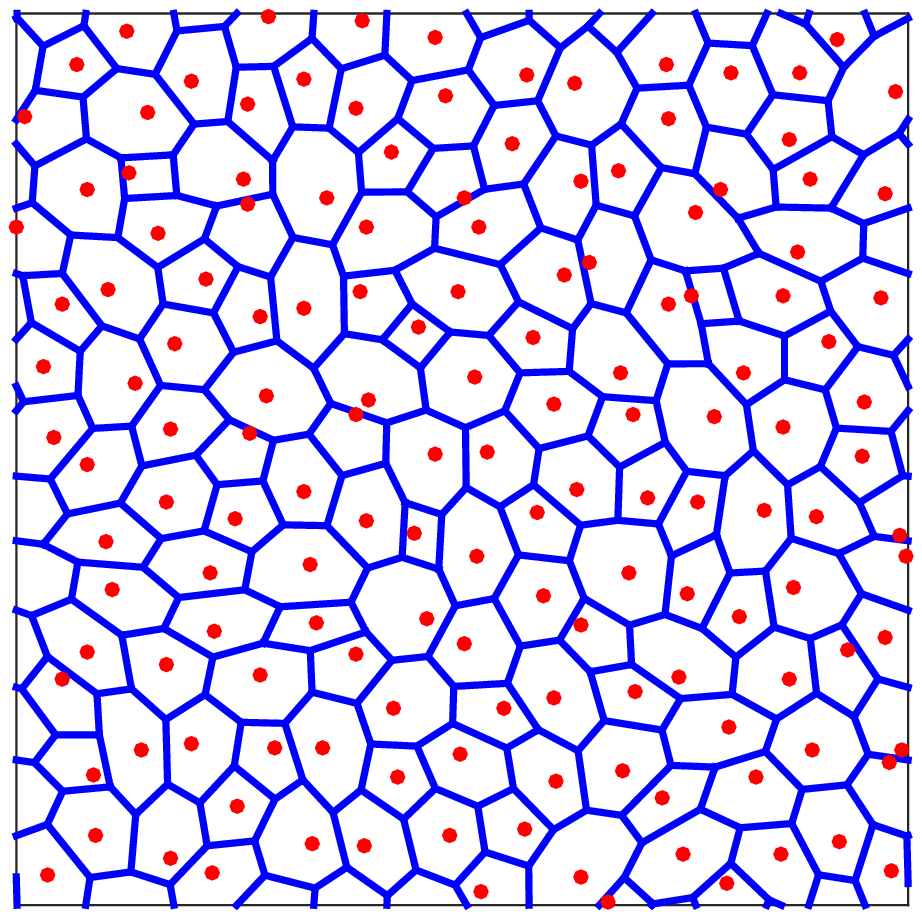} \\
\mbox{\bf (a)} & \mbox{\bf (b)} & \mbox{\bf (c)} \\
\includegraphics[width=0.32\textwidth]{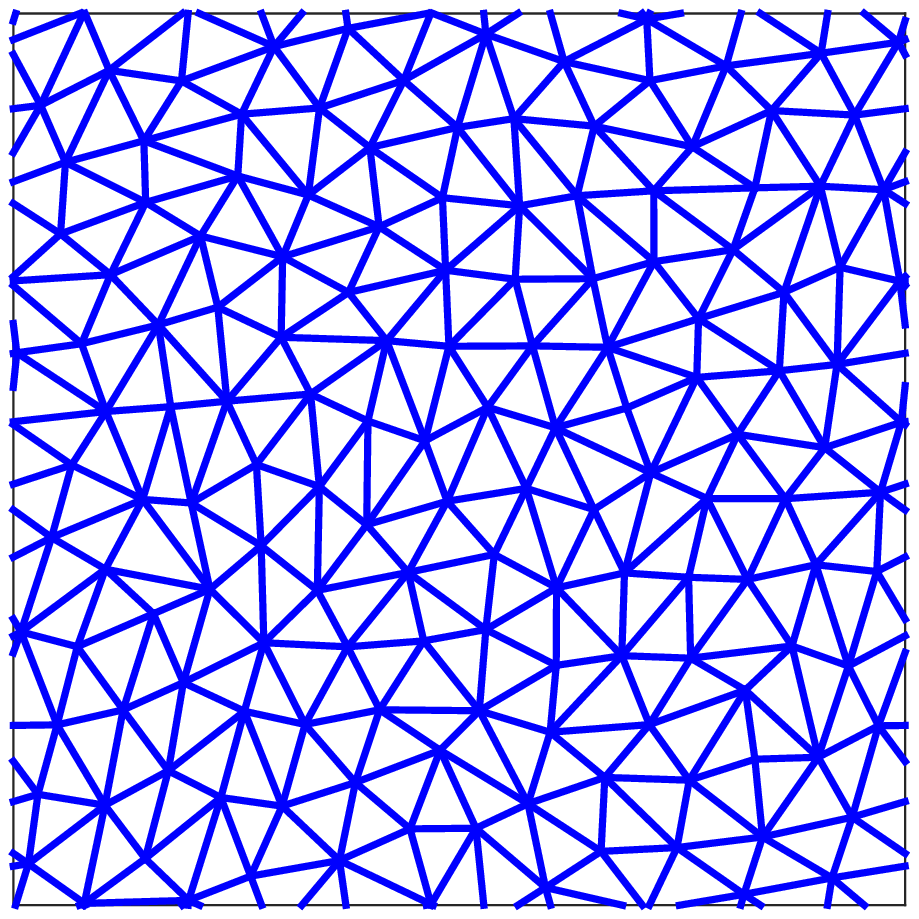} &
\includegraphics[width=0.32\textwidth]{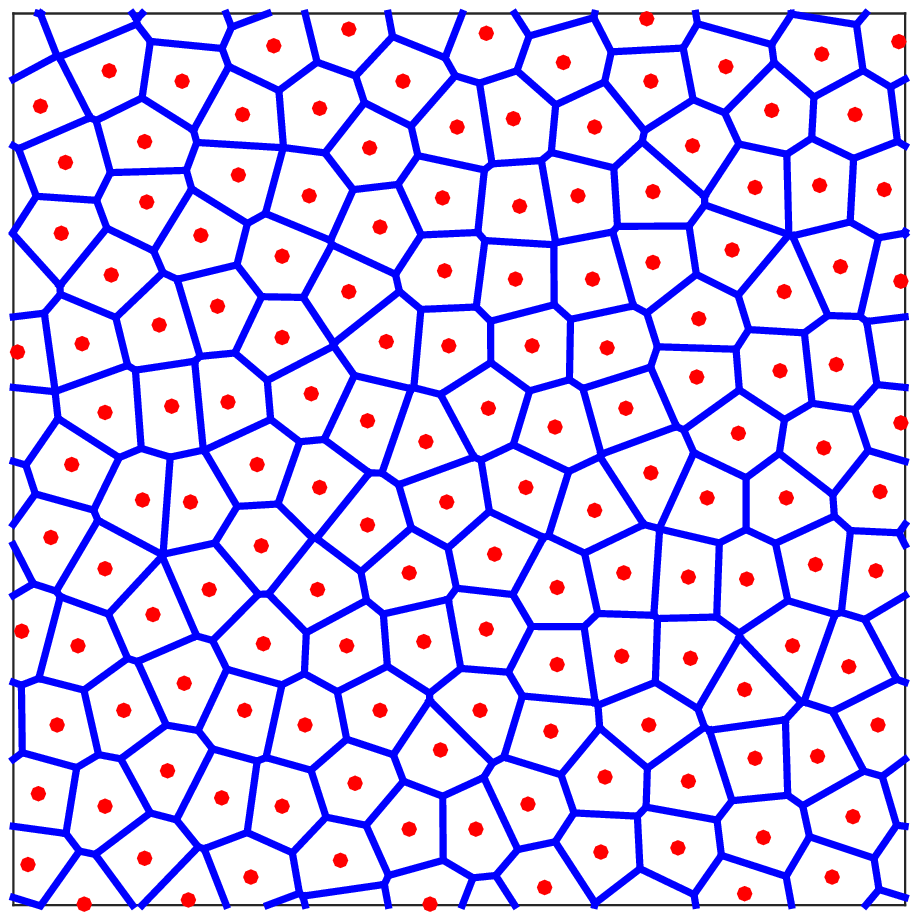} &
\includegraphics[width=0.32\textwidth]{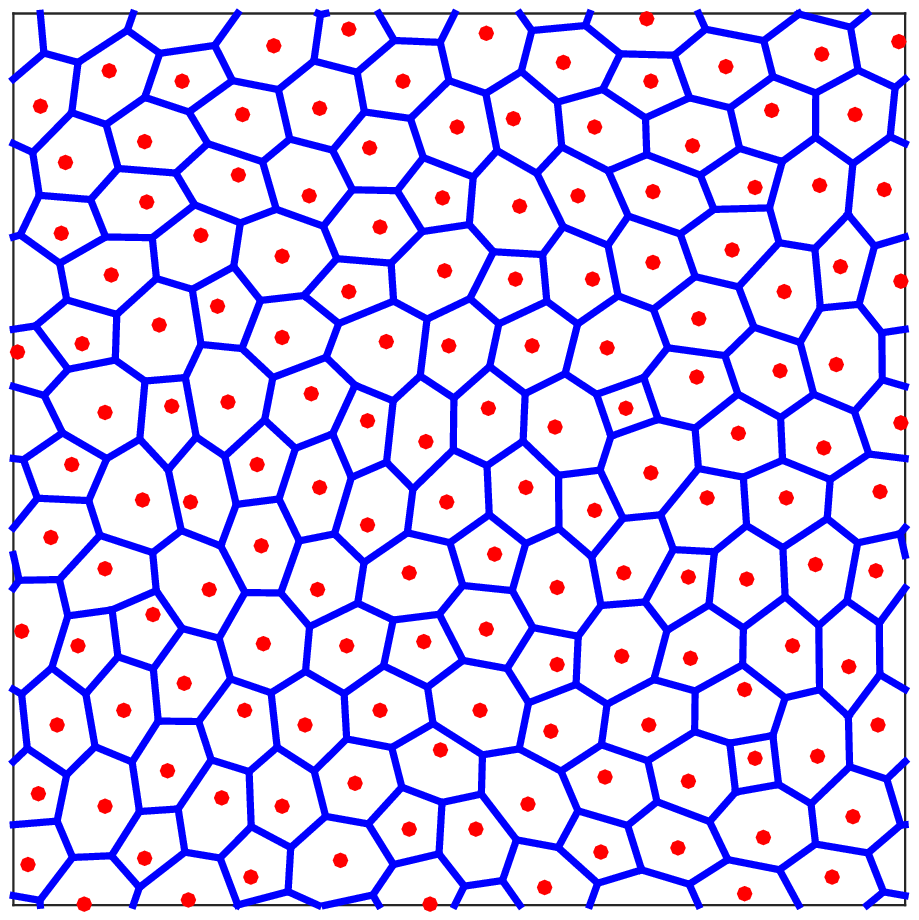} \\
\mbox{\bf (d)} & \mbox{\bf (e)} & \mbox{\bf (f)} \\
\includegraphics[width=0.32\textwidth]{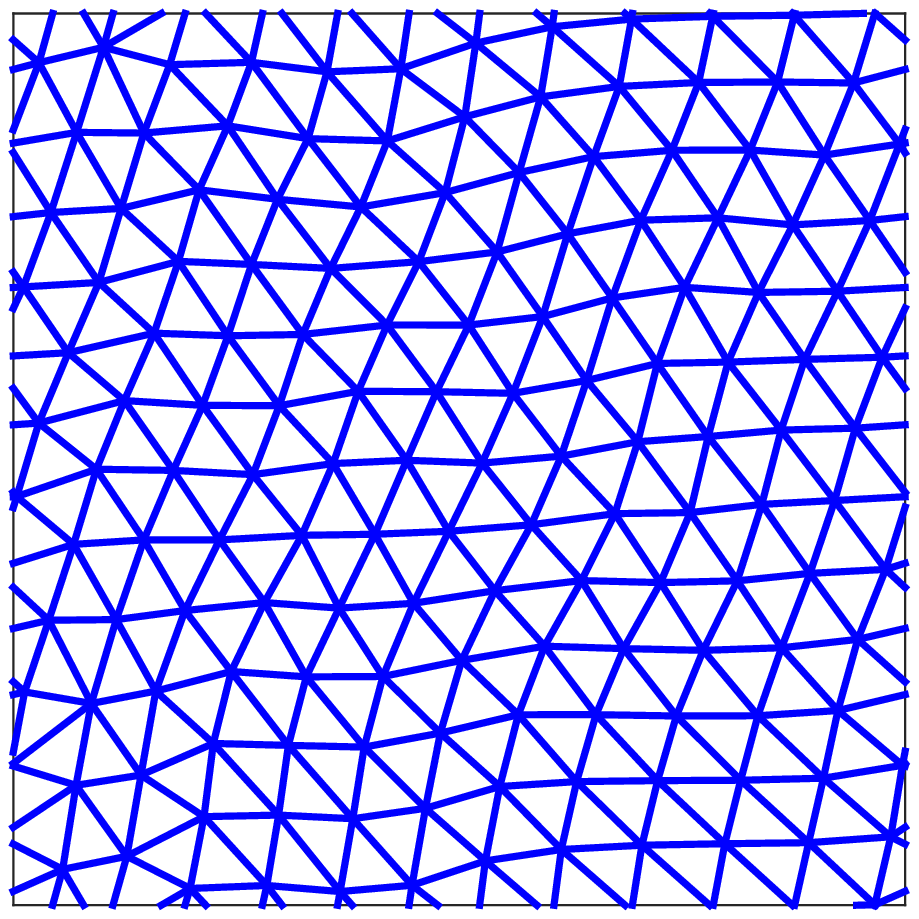} &
\includegraphics[width=0.32\textwidth]{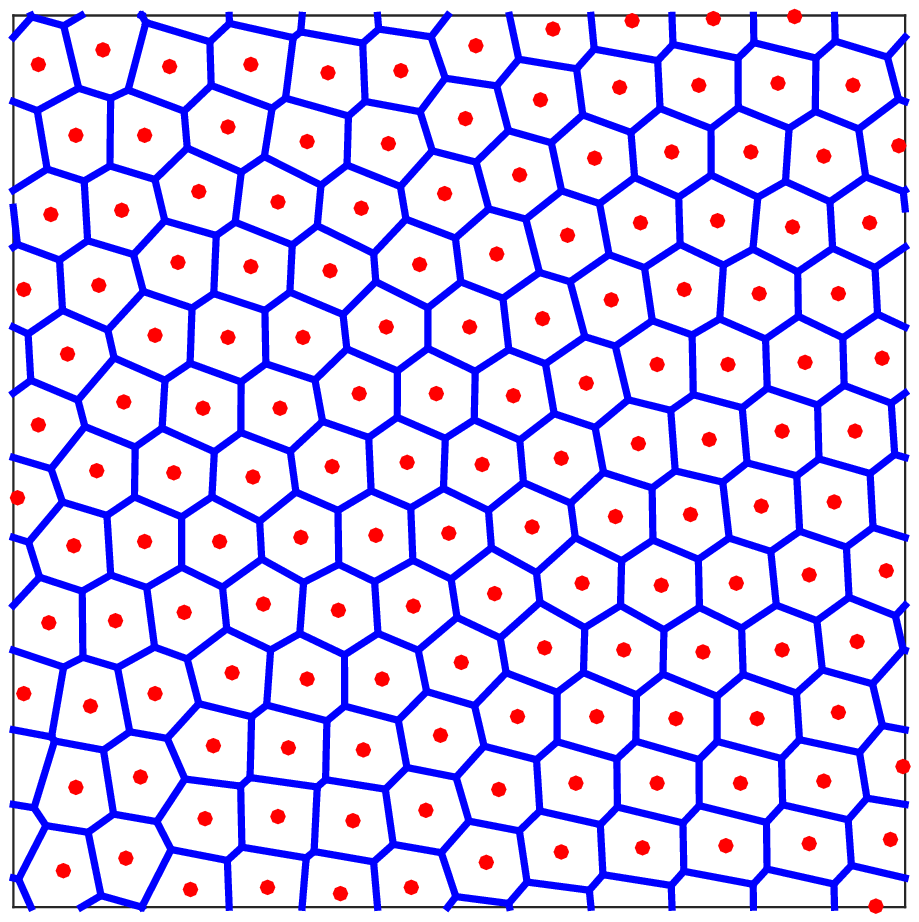} &
\includegraphics[width=0.32\textwidth]{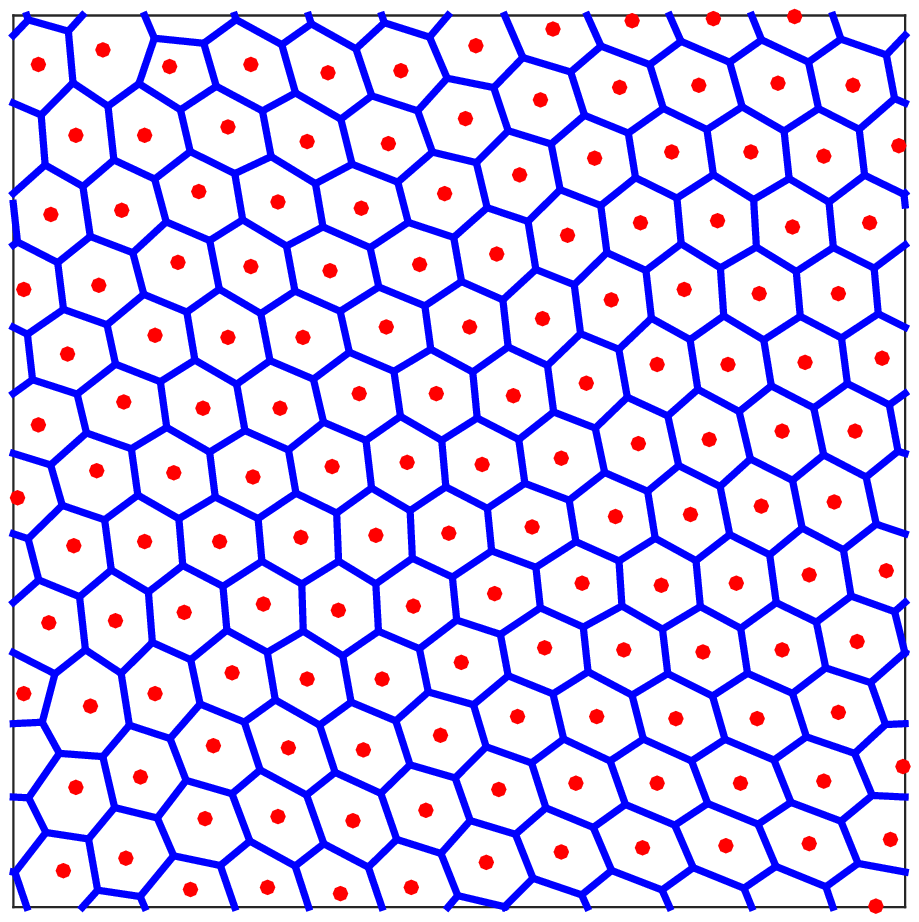} \\
\mbox{\bf (g)} & \mbox{\bf (h)} & \mbox{\bf (i)} \\
\end{array}$
\end{center}
\caption{Representative disordered stealthy cellular network structures mapped from various point patterns. There are $N = 150$ points in each underlying point pattern. Note that in those Voronoi and Delaunay-centroidal networks,
the underlying point patterns are colored in red, and the conducting ``channels'' are colored in blue. In those Delaunay
networks, the points in the underlying point patterns are just the vertices of the triangles, which are colored in blue.
(a) Delaunay network of stealthy point pattern with $\chi=0.3$. (b) Voronoi network of stealthy point pattern with $\chi=0.3$.
(c) Delaunay-centroidal network of stealthy point pattern with $\chi=0.3$. (d) Delaunay network of stealthy point pattern with $\chi=0.4$.
(e) Voronoi network of stealthy point pattern with $\chi=0.4$. (f) Delaunay-centroidal network of stealthy point pattern with $\chi=0.4$.
(g) Delaunay network of stealthy point pattern with $\chi=0.49$. (h) Voronoi network of stealthy point pattern with $\chi=0.49$.
(i) Delaunay-centroidal network of stealthy point pattern with $\chi=0.49$.} \label{fig_6}
\end{figure*}

The three types of constructed cellular network 
structures corresponding to Poisson, RSA and MRJ
point patterns that are not stealthy are shown in Fig. \ref{fig_5}, while those corresponding to disordered stealthy hyperuniform point patterns are shown in Fig. \ref{fig_6}. Note that in those Voronoi and Delaunay-centroidal networks, the underlying point patterns are colored in red, and the
conducting ``channels'' are colored in blue. In those Delaunay networks, the points in the underlying point patterns are just the vertices
 of the triangles, which are colored in blue. As $\chi$ increases, 
the fraction of hexagonal cells compared to all other 
possible polygonal cells in the corresponding networks increases, which is a manifestation 
of the increasing short-range order of the networks. Indeed, at $\chi = 0.49$, the average fraction
of hexagonal, pentagonal, and heptagonal cells for the Voronoi
and Delaunay-centroidal networks (averaged over ten
configurations) is equal to 96.8$\%$, 1.6$\%$ and 1.6$\%$,
respectively. Observe that all the cellular network structures considered here are statistically isotropic by construction,
and hence their effective conductivity is a scalar, which we denote by $\sigma_e$. 
Moreover, we conjecture that the networks derived from the stealthy hyperuniform point patterns are also stealthy and hyperuniform, which is based on strong numerical 
evidence from a previous photonic study \cite{Ma13}. However, 
we note that in a rigorous mathematical sense, this is still an open 
question, as we discuss in Sec. 7.


\subsection{Effective Conductivity}

Here we compute the effective conductivity $\sigma_e$ and 
tortuosity $\tau$ of these disordered statistically isotropic network structures by computationally solving
the equation described in Sec. 5.1. For each system, we average over ten configurations. The results are summarized in Table \ref{table_3}.

\begin{table*}[ht!]
\caption{Effective conductivity $\sigma_e$ (scaled by the conductivity $\sigma$ and volume fraction $\phi$ of the conducting ``channels'') and tortuosity $\tau$ of various isotropic disordered hyperuniform
and nonhyperuniform networks. The results are averaged over ten configurations for each system.}
\begin{center}
\begin{tabular}{{c}{c}{c}{c}} \\ \hline\hline
Point pattern & Tessellation & $\sigma_{e}/(\sigma\phi)$ & $\tau$ \\
\hline
Poisson & Delaunay & 0.4643 & 1.0769 \\
\hline
Poisson & Voronoi & 0.4473 & 1.1178 \\
\hline
Poisson & Delaunay-centroidal & 0.4490 & 1.1136 \\
\hline
RSA & Delaunay & 0.4860 & 1.0288 \\
\hline
RSA & Voronoi & 0.4907 & 1.0190 \\
\hline
RSA & Delaunay-centroidal & 0.4859 & 1.0290 \\
\hline
MRJ & Delaunay & 0.4887 & 1.0231 \\
\hline
MRJ & Voronoi & 0.4971 & 1.0058 \\
\hline
MRJ & Delaunay-centroidal & 0.4890 & 1.0225 \\
\hline
Stealthy with $\chi=0.3$ & Delaunay & 0.4698 & 1.0643 \\
\hline
Stealthy with $\chi=0.3$ & Voronoi & 0.4842 & 1.0326 \\
\hline
Stealthy with $\chi=0.3$ & Delaunay-centroidal & 0.4677 & 1.0691 \\
\hline
Stealthy with $\chi=0.4$ & Delaunay & 0.4751 & 1.0524 \\
\hline
Stealthy with $\chi=0.4$ & Voronoi & 0.4896 & 1.0212 \\
\hline
Stealthy with $\chi=0.4$ & Delaunay-centroidal & 0.4762 & 1.0500 \\
\hline
Stealthy with $\chi=0.49$ & Delaunay & 0.4937 & 1.0128 \\
\hline
Stealthy with $\chi=0.49$ & Voronoi & 0.4984 & 1.0032 \\
\hline
Stealthy with $\chi=0.49$ & Delaunay-centroidal & 0.4952 & 1.0097 \\
\hline\hline
\end{tabular}
\end{center}
\label{table_3}
\end{table*}

It is noteworthy that Poisson networks have the lowest effective conductivity due to the complete absence of order on all length scales. On the other hand, for those point patterns associated with hard-disk packings, as the packings approach jamming
and the point patterns tend toward hyperuniform states, the effective conductivity of the corresponding network structures
increases. Moreover, for those point patterns that are indeed hyperuniform and stealthy, as $\chi$ increases, i.e., the short-range
order of the corresponding networks dramatically increases \cite{Zh15}, the effective conductivity 
of the corresponding network increases. Interestingly, when $\chi = 0.49$,
the corresponding statistically isotropic networks are nearly optimal in terms of their effective conductivity, i.e., achieve the upper bound (\ref{eq_53}). 
These observations suggest that for disordered statistically isotropic Voronoi, 
Delaunay, and ``Delaunay-centroidal'' cellular 
network structures to achieve optimal effective conductivity, both short-range and long-range orders are necessary. These networks are ideal for 
heat dissipation as well as electrical and fluid
(see Sec. 7) transport through the channel phase. In addition, among the three types of 
tessellations investigated here, the Voronoi tessellations generally 
possess higher effective conductivity than
the Delaunay and Delaunay-centroidal tessellations of the same point pattern, 
except for the Poisson point pattern.

\clearpage
\subsection{Cross-Property Relations}

The disordered networks investigated here are elastically isotropic and hence are characterized by two independent moduli. Here we compute the upper bounds on the effective bulk moduli of these structures using
the cross-property bound (\ref{eq_40}) and the shear moduli 
of the Delaunay networks using the bounds (\ref{eq_41}) 
and (\ref{eq_42}). The results are summarized in Table \ref{table_4}. It is noteworthy 
that similar to the conduction problem, as both short-range and long-range order of the 
network increases, the effective bulk moduli of all the networks and the shear moduli of the Delaunay networks increase. 
Specifically, when $\chi = 0.49$, the corresponding statistically isotropic stealthy
networks possess nearly optimal effective bulk moduli, and the Delaunay ones among them possess nearly optimal effective shear moduli as well.

\begin{table*}[ht!]
\caption{Upper bounds on the effective moduli $K_e$, $G_e$, and $E_e$ of the various isotropic disordered networks summarized in Table \ref{table_3} and shown in Figs. \ref{fig_5} and \ref{fig_6}. 
Here the effective properties are scaled by the corresponding moduli $K$, $G$, $E$, and volume fraction $\phi$ of the ``channel'' phase. The results are averaged over ten configurations for each system. The results for $G_e$ and $E_e$ are only shown for Delaunay networks. Note that disordered 
statistically isotropic stealthy cellular 
networks with $\chi=0.49$ possess nearly optimal bulk moduli.}
\begin{center}
\begin{tabular}{{c}{c}{c}{c}{c}} \\ \hline\hline
Point pattern & Tessellation & $K_{e}/(K\phi)$ & $G_{e}/(G\phi)$ & $E_{e}/(E\phi)$ \\
\hline
Poisson & Delaunay & $0.4643(1-\nu)$~~ & $0.2322(1+\nu)$ & 0.3095 \\
\hline
Poisson & Voronoi & $0.4473(1-\nu)$~~ & &  \\
\hline
Poisson & Delaunay-centroidal & $0.4490(1-\nu)$~~ & & \\
\hline
RSA & Delaunay & $0.4860(1-\nu)$~~ & $0.2430(1+\nu)$ & 0.3240 \\
\hline
RSA & Voronoi & $0.4907(1-\nu)$~~ & & \\
\hline
RSA & Delaunay-centroidal & $0.4859(1-\nu)$~~ & & \\
\hline
MRJ & Delaunay & $0.4887(1-\nu)$~~ & $0.2444(1+\nu)$ & 0.3258 \\
\hline
MRJ & Voronoi & $0.4971(1-\nu)$~~ & & \\
\hline
MRJ & Delaunay-centroidal & $0.4890(1-\nu)$~~ & & \\
\hline
Stealthy with $\chi=0.3$ & Delaunay & $0.4698(1-\nu)$~~ & $0.2349(1+\nu)$ & 0.3132 \\
\hline
Stealthy with $\chi=0.3$ & Voronoi & $0.4842(1-\nu)$~~ & & \\
\hline
Stealthy with $\chi=0.3$ & Delaunay-centroidal & $0.4677(1-\nu)$~~ & & \\
\hline
Stealthy with $\chi=0.4$ & Delaunay & $0.4751(1-\nu)$~~ & $0.2376(1+\nu)$ & 0.3167 \\
\hline
Stealthy with $\chi=0.4$ & Voronoi & $0.4896(1-\nu)$~~ & & \\
\hline
Stealthy with $\chi=0.4$ & Delaunay-centroidal & $0.4762(1-\nu)$~~ & & \\
\hline
Stealthy with $\chi=0.49$ & Delaunay & $0.4937(1-\nu)$~~ & $0.2469(1+\nu)$ & 0.3291 \\
\hline
Stealthy with $\chi=0.49$ & Voronoi & $0.4984(1-\nu)$~~ & & \\
\hline
Stealthy with $\chi=0.49$ & Delaunay-centroidal & $0.4952(1-\nu)$~~ & & \\
\hline\hline
\end{tabular}
\end{center}
\label{table_4}
\end{table*}
\vspace{0.5in}
\newpage
\section{Conclusion and Discussion}

In this work, we considered and constructed various 2D ordered 
and disordered low-density cellular networks, and 
determined their effective conductivities, tortuosity tensors, and elastic moduli.
In particular, we investigated periodic hyperuniform networks including both macroscopically isotropic and anisotropic varieties, as well as 
various disordered statistically isotropic networks derived from Voronoi, Delaunay, and
``Delaunay-centroidal'' tessellations based on 
hyperuniform and nonhyperuniform point patterns. 
We observed that the presence of ``dead ends'' 
in a network leads to suboptimal effective conductivity.
We also demonstrated for the first time that intersecting parallel-channel 
cellular networks, including disordered hyperuniform and nonhyperuniform varieties, 
possess optimal effective conductivity tensors.
We find that the effective conductivities and elastic moduli of the disordered 
Voronoi, Delaunay, and ``Delaunay-centroidal'' networks 
correlated positively with the short-range and long-range order of the networks, which is consistent with the fact that Poisson networks have the lowest effective properties due to the absence of any order.
Moreover, we found that certain disordered networks 
derived from disordered stealthy hyperuniform 
point patterns with $\chi$ values just below $1/2$ maximize 
heat (or electrical) conduction/dissipation and fluid transport through the solid phase, and are 
capable of sustaining external stress with minimal amount of deformation. 
The Delaunay ones among them possess nearly 
optimal effective shear moduli as well. In summary, 
the effective transport and elastic properties of 
disordered networks derived from stealthy point patterns generally improve
as the short-range order
increases due to an increasing value of $\chi$ within the disordered 
regime ($\chi<1/2$). This is also supported
by a previous study \cite{Fl09}
in which the size of the photonic band gap of a disordered stealthy
hyperunuform dielectric network was
shown to be proportional
to $\chi$.

It should not go unnoticed that
all of the results that we have obtained for the effective conductivity apply as well
to the fluid permeability associated with slow viscous flow through the channels.
This is because the Stokes-flow
equations for fluid transport
in networks  in the low-density
limit ($\phi \to 0$) become identical
to the conduction governing equations \cite{To02a}.
Thus, networks that are optimal with respect 
to the effective conductivity 
are also optimal with respect
to the fluid permeability. Moreover, because
the fluid permeability has been shown
to be directly linked to the mean survival
time associated with diffusion-controlled reactions in channels 
\cite{To90}, our results for the effective conductivity are also optimal
for the mean survival time.

The variety of favorable properties 
make these low-weight networks ideal for applications that 
require multifunctionality with respect
to transport, mechanical and electromagnetic properties, 
e.g., aerospace applications \cite{No00}. 
Such low-weight multifunctional networks can be 
readily fabricated using 3D printing and lithographic technologies \cite{Va13,Co14}. 
In addition, although the procedures and 
results in this work focused on two dimensions, they can be easily extended to treat 
three-dimensional open-cell foams, where the void phase is interconnected, which may 
have potential biomedical applications \cite{Hu00}.

While the identified optimal
networks were derived
in the low-density limit ($\phi \to 0$), we expect that they
remain optimal for small
but positive volume fractions
and may even apply at
intermediate values of $\phi$
when the channels are
``thickened." Previous work
described in Refs. \cite{Hy00,Hy02}
supports this conjecture.
Confirming this conjecture 
represents a worthy
subject for future research.

It is useful to note that disordered networks derived from 
disordered hyperuniform point patterns are not necessarily hyperuniform. 
This is related to the fact that the centroids of the polygons 
in the disordered network do not necessarily coincide with the points in the 
disordered point pattern that is used to generate the network \cite{Kl18}. 
A previous 
numerical study of dielectric networks
derived from stealthy point configurations \cite{Ma13} strongly suggests that 
these networks
are also stealthy and hyperuniform.
However, the rigorous mathematical conditions 
required to transform stealthy hyperuniform point
patterns into stealthy 
hyperuniform networks 
have yet to be identified.
By contrast, ordered networks derived from ordered hyperuniform
point patterns are always hyperuniform. For example, the honeycomb network 
associated with the Voronoi tessellations of the hyperuniform point pattern of 
triangular lattice is hyperuniform. Moreover, 
the spectral density of the honeycomb network 
$[\widetilde{\chi}_{_V}({\bf k})]_H$ is proportional to the structure 
factor of the triangular lattice $[S({\bf k})]_T$, i.e., 
\begin{equation}
\label{eq_70} [\widetilde{\chi}_{_V}({\bf k})]_H=\rho\widetilde{m}^2({\bf k})[S({\bf k})]_T,
\end{equation}
where $\rho$ is the number density of the triangular lattice, and $\widetilde{m}({\bf k})$
is the Fourier transform of the indicator function of the material in the fundamental cell 
(the smallest repeating hexagonal unit) 
of the honeycomb network. 
The investigation of the relationship between the hyperuniformity 
of disordered point patterns and the hyperuniformity of the generated disordered network could 
shed light on identifying novel ways to generate disordered hyperuniform networks. 
\section*{Acknowledgments}
The authors gratefully acknowledge the support
of the Air Force Office of
Scientific Research Program
on Mechanics of Multifunctional
Materials and Microsystems under Award No. FA9550-18-1-0514.

\clearpage
\section*{References}

\providecommand{\newblock}{}

\end{document}